\documentclass[preprint,english,showpacs,11pt,floatfix]{revtex4}

\usepackage{babel,amsmath,amssymb,dcolumn}
\usepackage[dvips]{graphics}
\usepackage{hyperref}

\usepackage{amsfonts}
\usepackage{amssymb}
\usepackage[dvips]{graphicx}
\usepackage{latexsym}
\usepackage[nooneline]{caption2}
\usepackage{subfigure}

\usepackage{dcolumn}
\usepackage{bm}

\begin{document}

\title{Uncertainty on determining the dark energy equation of state due to the spatial curvature}

\author{Zhuo-Yi Huang, Bin Wang}
\email{wangb@fudan.edu.cn} \affiliation{Department of Physics,
Fudan University, Shanghai 200433, People's Republic of China }

\author{Ru-Keng Su}
\email{rksu@fudan.ac.cn} \affiliation{China Center of Advanced
Science and Technology (World Laboratory), P.B.Box 8730, Beijing
100080, People's Republic of China
\\Department of Physics, Fudan University, Shanghai 200433,
People's Republic of China }

\begin{abstract}
We have studied the uncertainty on the determination of the dark
energy equation of state due to a non-vanishing spatial curvature by
considering some fundamental observables. We discussed the
sensitivity of these observables to the value and redshift history
of the equation of state and the spatial curvature and investigated
whether these different observables are complementary and can help
to reduce the cosmic confusion.
\end{abstract}


\maketitle

There is growing observational evidence indicating that our universe
is undergoing an accelerated expansion driven by a yet unknown dark
energy (DE) \cite{01}. In the past years much effort has been made
to understand the nature and the origin of the DE. The leading
interpretation of such a DE is a cosmological constant with equation
of state (EoS) $w_D=-1$. There are other conjectures relating the DE
to a scalar field called Quintessence with $w_D>-1$, or to an exotic
field called Phantom with $w_D<-1$ \cite{02}. Recently, extensive
analysis found that the current data favors DE models with EoS in
the vicinity of $w_D=-1$ \cite{03}, straddling the cosmological
constant boundary. Since the EoS can give insight into the
microscopic nature of DE, it is of particular interest to
investigate whether the observations will allow us to determine
formally if DE is of the Phantom type, Quintessence type or
cosmological constant.

In observations, nearly all proposed tests for the nature of DE
measure some combination of four fundamental observables: the Hubble
parameter $H(z)$, the distance-redshift relation $d_L(z)$, the
age-redshift relation $t(z)$, or the linear growth factor $D_1(z)$
\cite{04,05}. Values of these observables are sensitive to redshift
and the EoS. Studying DE tests in terms of these observables gives
us advantage to see whether and how different observational
strategies complement each other so that can help to constrain the
DE EoS. In \cite{04}, sensitivities of measurements of different
observables or of the same observable at different redshifts for
constant EoS or dynamic EoS have been discussed and whether these
different observables are complementary has been emphasized. Their
work was done based on the assumption that the universe is flat and
has not included the spatial curvature influence.

Usually, investigations of DE assume a spatially flat universe.
However tendency of preferring a closed universe appeared in a suite
of CMB experiments \cite{06}. The improved precision from WMAP
provides further confidence, showing that a closed universe with
positively curved space is marginally preferred \cite{07}. In
addition to CMB, recently the spatial geometry of the universe was
probed by supernova measurements of the cubic correction to the
luminosity distance \cite{08}, where a closed universe is also
marginally favored. Looking towards future precision of
observational data, it is worthwhile examining DE in some detail
for a non-flat space. Some attempts at the investigation
of DE in the universe with spatial curvature have been carried out
\cite{09,10,11}. In \cite{09,10}, it was found that a non-flat
$\Lambda$ universe is compatible with the SN+CMB data. The extension
of quantifying the parameter space $(w,\Omega_{k}^0)$ for the static
DE with constant EoS was done in \cite{11}. It was shown from the
measurement of the luminosity distance that there is significant
uncertainty on the determination of DE EoS due to a non-vanishing
spatial curvature. It was argued that this kind of cosmic confusion
could be reduced by comparing the luminosity distance measurements
at different redshifts.

The motivation of the present paper is to extend the study of the
cosmic confusion caused by the non-vanishing spatial curvature by
considering the DE with dynamic EoS. We will choose two
time-dependent DE models with different parameterizations:
\begin{eqnarray}
  w_D^\textrm{I}(z) &=& w_0+w_1\frac{z}{1+z} \label{para1}, \\
  w_D^{\textrm{II}}(z) &=& w_0+w_1\frac{z}{(1+z)^2}. \label{para2}
\end{eqnarray}
These two models have been extensively discussed in various papers,
for example, see \cite{12,13,14,15,16}. Besides concentrating on the
measurements of the luminosity distances as done in \cite{11}, we
will also examine other fundamental observables, such as the Hubble
parameter $H(z)$, the linear growth factor $D_1(z)$ and the
Alcock-Pacyznski parameter $d_L(z)H(z)$ \cite{17}. We will
investigate whether different observables are complementary and
helpful to reduce the cosmic confusion brought by the spatial
curvature and the time-dependent DE EoS.

Let us first go over definitions of different observables. The
accelerated expanding universe is described by the Friedmann
euqation,
\begin{equation}\label{Firedmann}
    H^2(z)=H_0^2 \left[ \Omega_m^0(1+z)^3 + \Omega_D^0 f(z) + \Omega_k^0(1+z)^2 \right],
\end{equation}
where $\Omega_m^0$, $\Omega_D^0$ and $\Omega_k^0$ refer to densities
 of matter, dark energy and curvature at the present day in units of the
 critical density. The function
$f(z)$ is related to the EoS of DE by
\begin{equation}\label{fz}
    f(z) = \exp \left[ 3 \int_0^z \frac{1+w(z')}{1+z'}\, dz' \right].
\end{equation}

Besides the Hubble parameter described in Eq. (\ref{Firedmann}), there
are other fundamental observables such as the luminosity distance
$d_L(z)$ and the linear growth factor of mass fluctuation $D_1(z)$. The
luminosity distance $d_L(z)$ can be expressed as
\begin{equation}\label{dL}
    d_L(z) = \frac{1+z}{H_0} \frac{1}{\sqrt{|\Omega_k^0|}}\, S\left(
    \sqrt{|\Omega_k^0|} \int_0^z \frac{dz'}{h(z')} \right),
\end{equation}
where $h(z)=H(z)/H_0$ and the function $S(x)$ takes the form
$\sin(x)$, $x$ and $\sinh(x)$ for a closed, flat and open universe,
respectively. The linear growth factor is the solution to the
differential equation,
\begin{equation}\label{D1eq}
    \ddot{D_1} + 2\,H(z)\,\dot{D_1} - \frac{3}{2}\, \Omega_m\, H_0^2\,(1+z)^3
    D_1 =0
\end{equation}
The approximation $d\log D1/d\log a \equiv f(\Omega_m)\approx
\Omega_m^{4/7}$ leads the solution of Eq.\ref{D1eq} to \cite{05}
\begin{equation}\label{D1}
    D_1(z) = \exp \left[ - \int_0^z \frac{dz'}{1+z'} \left( \frac{\Omega_m(1+z')^3}{h(z')} \right)^{4/7}\right].
\end{equation}
There is another interesting observable called the Alcock-Pacyznski
(AP) parameter, which is the product $d_L(z)H(z)$.

We know that different observations measure different observables.
For example, studies of Type IA supernovae measure $d_L(z)$
directly, while the observations of weak lensing are sensitive to
$D_1(z)$, $d_L(z)$ and $H(z)$. The AP parameter is useful in the
Alcock-Pacyznski anisotropy test \cite{17}. Discussing DE tests in
terms of different fundamental observables can show us how different
observational strategies complement each other and help to reduce
the comic confusion.

\begin{figure}[!hbtp]
  \centering
    \subfigure[ \hspace{0.1cm} Iso-$d_L$ contours with constant
    EoS ($w_1=0$) through $(\Omega_k=0,w_0=-1)$. Solid line is measured at $z=3.1$,
    dashed line at $z=0.2$ and dot-dashed line at $z=4$.]
    {\includegraphics[width=0.4\textwidth]{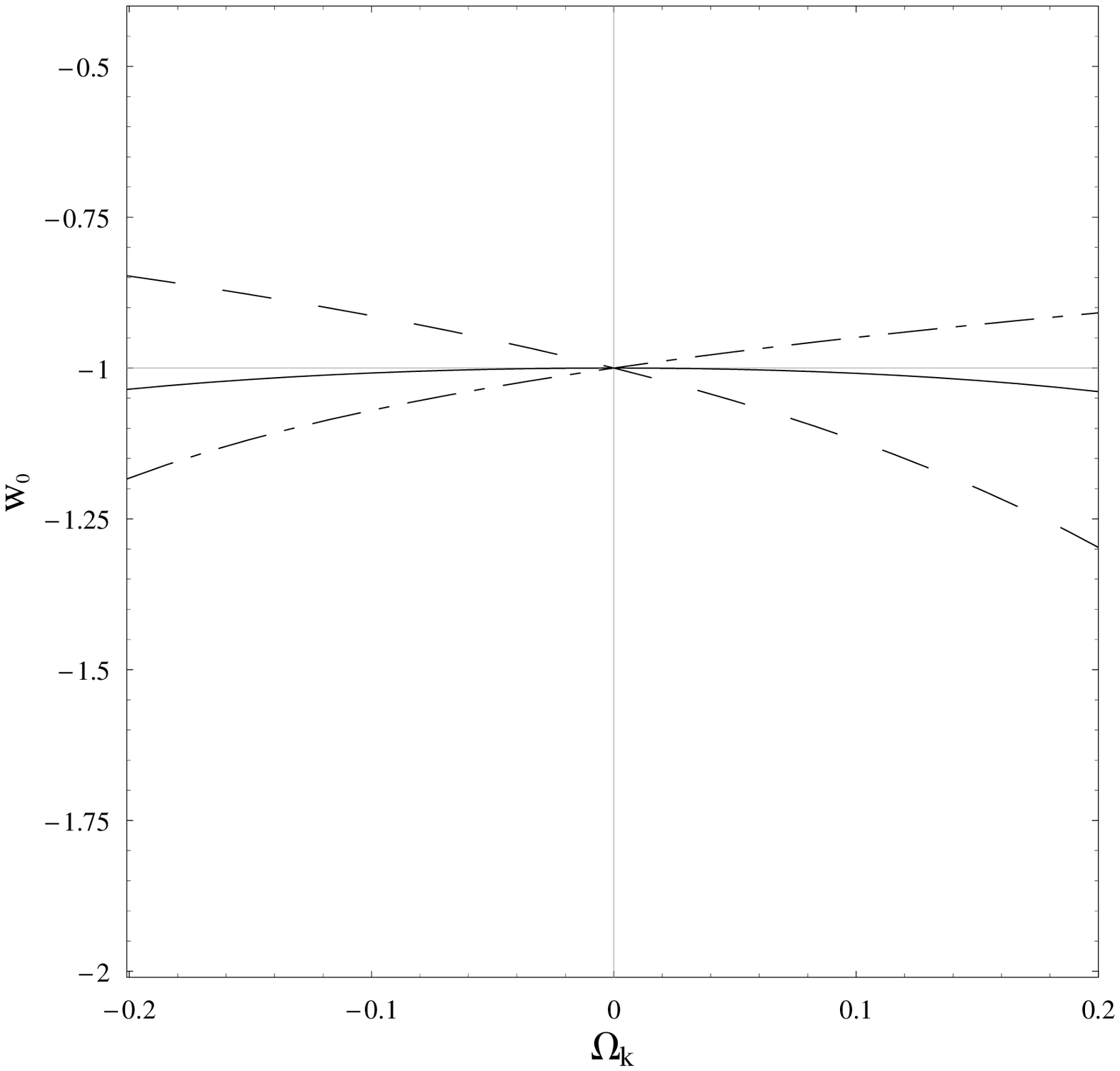}\label{fig1a}}
    \hspace{0.01\textwidth}
    \subfigure[ \hspace{0.1cm} For the time-dependent EoS, the critical redshift to get the minimal
degeneracy at $w_0=-1$ and its relation to $w_1$.]
    {\includegraphics[width=0.4\textwidth]{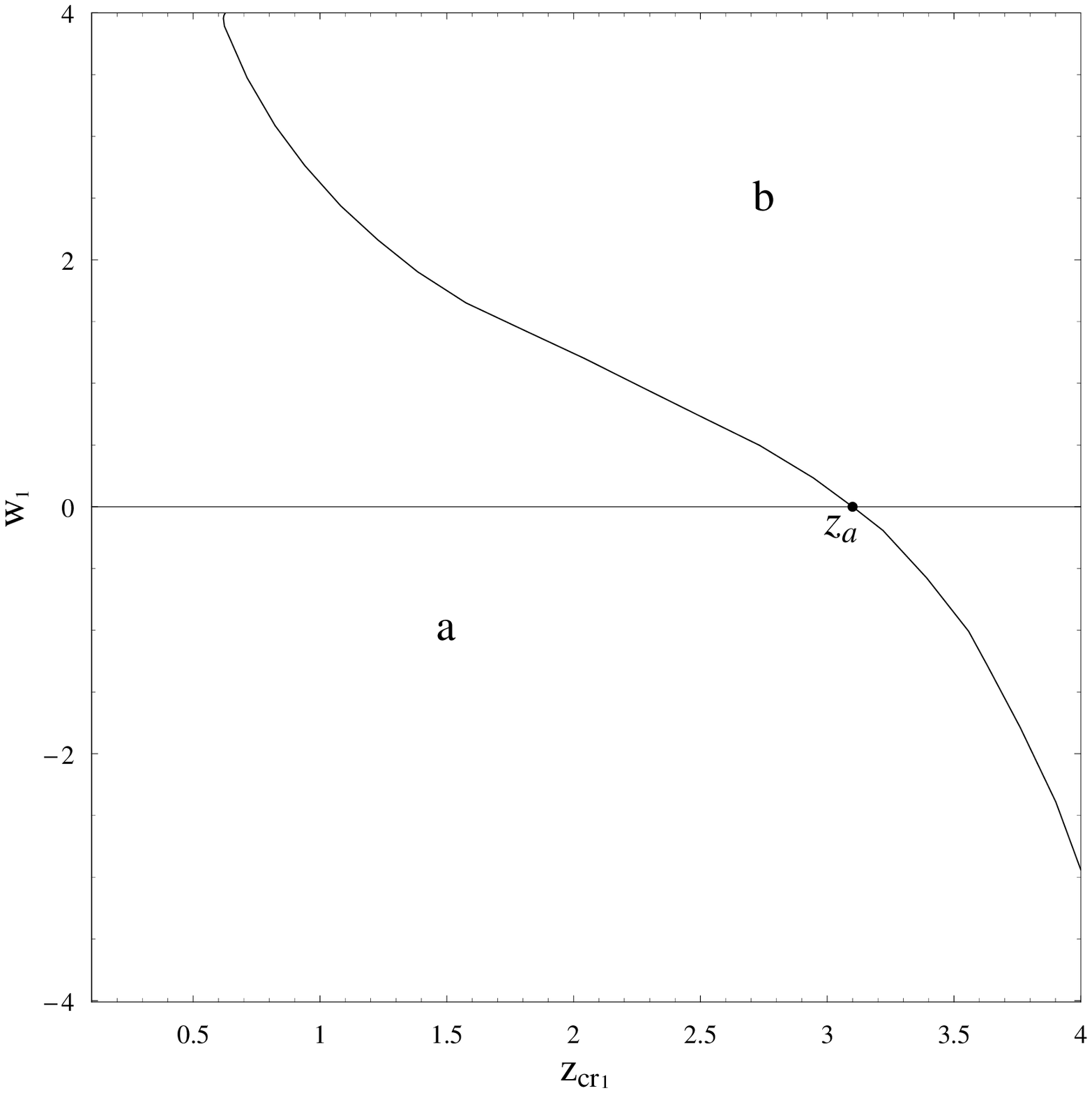}\label{fig1b}}
    \caption{.}\label{fig1}
\end{figure}

The cosmic confusion which arises due to a non-vanishing curvature
of the universe by considering the luminosity distance $d_L$ in the
two dimensional parameter space $(\Omega_k^0,w)$ for fixed
$\Omega_m^0$ and arbitrary constant $w$ has been found in \cite{11}.
It was shown that at low redshift $z_1$, the iso-$d_L$ curves
(having constant $d_L$ values) display degeneracy in interpreting
the observations: a pure cosmological constant $\Lambda$ assuming a
flat universe (Flat-$\Lambda$) can equally well be interpreted as
Quintessence DE with $w>-1$ if one takes closed universe (Closed-Q)
or Phantom DE with $w<-1$ if one takes open universe (Open-P) (see
dashed line in Fig.\ref{fig1a}). However at high redshift $z_2$, it
was found that the degeneracy in the $(\Omega_k^0,w)$ parameter
plane is opposite, the iso-$d_L$ curves display the cosmic confusion
among Flat-$\Lambda$, Closed-P and Open-Q, which is also shown in
the dot-dashed line in Fig.\ref{fig1a}. Comparing the property of
degeneracies for low and high redshifts, we can constrain the
degeneracy in $w$. Within the redshift range $[z_1,z_2]$, one can
find a critical redshift $z_a$, which exhibits no degeneracy in $w$
while maximal degeneracy with respect to the curvature as shown in
the solid line in Fig.\ref{fig1a}, when $\Omega_m^0$ is fixed at
0.3, $z_a=3.1$. At this critical redshift we expect to detect the
value of the constant EoS.

It is of great interest to extend the discussion to the
time-dependent EoS, since recent analysis indicates that the time
varying DE gives a better fit to the type Ia supernova data
\cite{18}, which mildly favors the evolution of the DE EoS from
$w>-1$ to $w<-1$ at a recent stage. Theoretical attempts towards
understanding the $w$ crossing $-1$ phenomenon have been taken
\cite{19}. In the following we will use two parameterization models
(\ref{para1}) (\ref{para2}) to investigate how a varying EoS would
influence the cosmic confusion. We will work now in the three
dimensional parameter space $(\Omega_k^0,w_0,w_1)$ for fixed
$\Omega_m^0$. We will explain in detail by using the
parameterization model (\ref{para1}). The result for
the second model (\ref{para2}) is similar.

We consider first the luminosity distance $d_L$ in our parameter
space $(\Omega_k^0,w_0,w_1)$ measured at fixed redshift $z$. For the
dynamic EoS, we see that for more positive $w_1$, we get much
smaller critical redshift $z_{cr1}$ where we have no degeneracy in
$w_0$ while maximal degeneracy with respect to the curvature. This
property is shown in Fig.\ref{fig1b}. At $w_1=0$, $z_{cr1}=z_a=3.1$.
The region  \textbf{a}  below the line in Fig.\ref{fig1b} contains
the cosmic degeneracy in observation (Closed-Q, Flat-$\Lambda$,
Open-P), while the region \textbf{b} above has the degeneracy among
(Closed-P, Flat-$\Lambda$, Open-Q).

\begin{figure}[!hbtp]
  \centering
    \subfigure[ \hspace{0.1cm} Iso-$d_L$ contours with $w_0=-1$ through $(\Omega_k=0,w_1=0)$. Solid line is got at $z=3.1$,
    dashed line at $z=0.2$ and dot-dashed line at $z=4$.]
    {\includegraphics[width=0.4\textwidth]{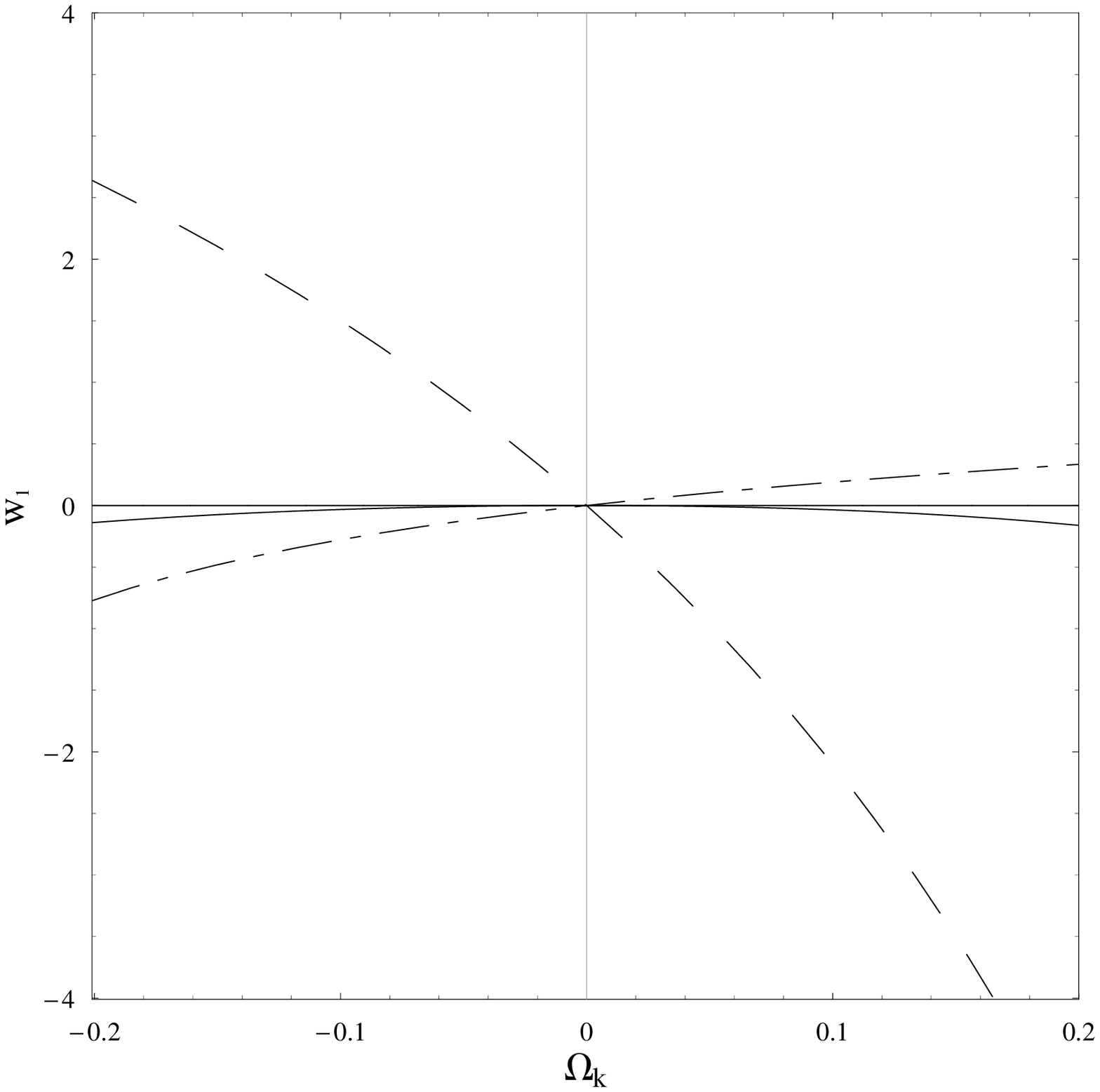}\label{fig2a}}
    \hspace{0.01\textwidth}
    \subfigure[  \hspace{0.1cm} The $w_0$-dependence of $z_{cr3}$, where minimal degeneracy at $w_1=0$. When $w_0=-1$,
    $z_a=3.1$. Region \textbf{a} and region
    \textbf{b} divided by the solid line have different cosmic degeneracy.]
    {\includegraphics[width=0.4\textwidth]{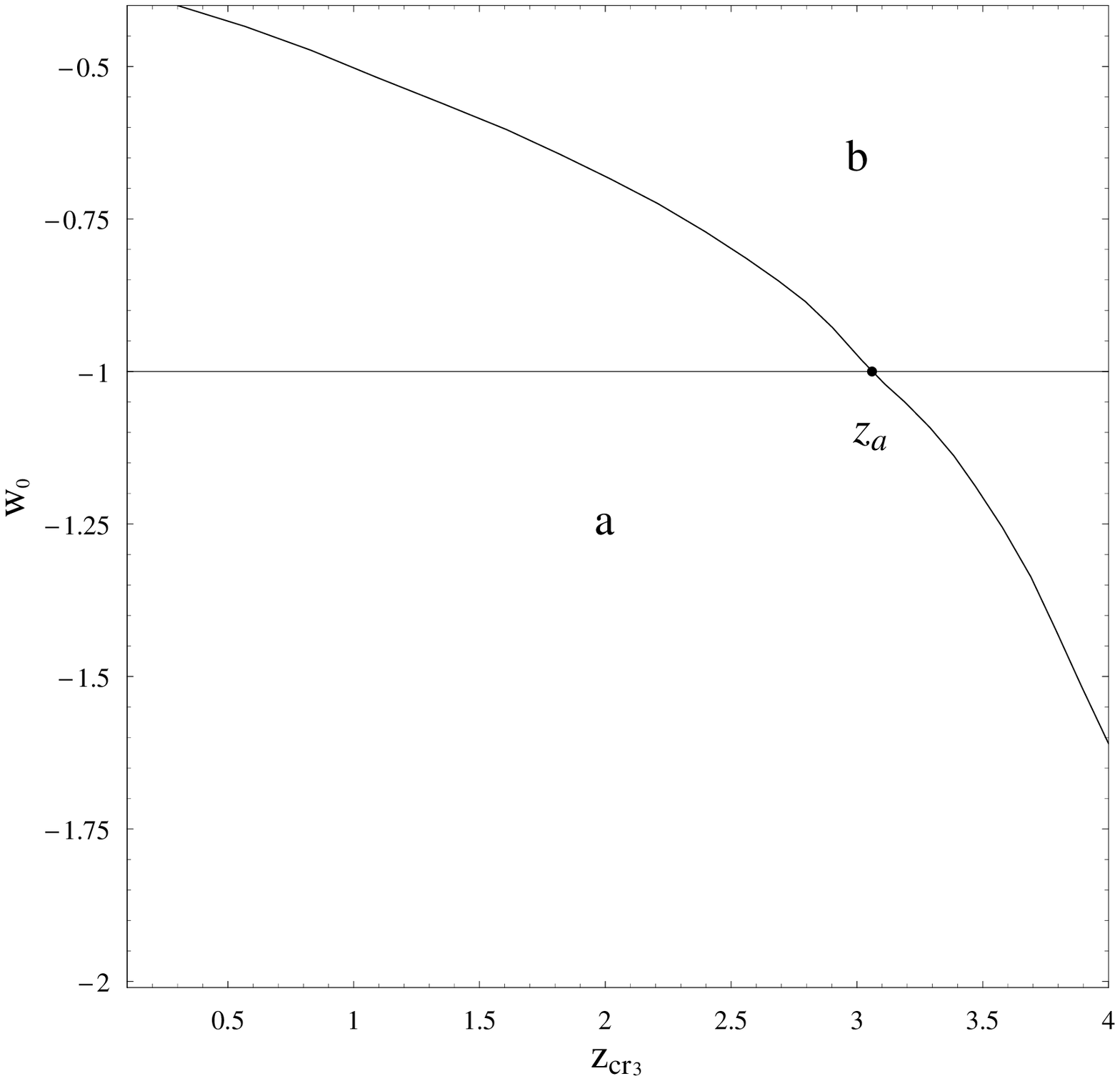}\label{fig2b}}
    \caption{.}\label{fig2}
\end{figure}

For fixed $w_0$, we have also studied the parameter space
$(\Omega_k^0,w_1)$ in measuring the $d_L(z)$. There exists a
critical redshift in studying the luminosity distance where there is
no degeneracy in $w_1$ while maximum degeneracy in $\Omega_k$ which
is shown by the solid line in Fig.\ref{fig2a}. This critical
redshift which we refer to as $z_{cr3}$ can help to determine
whether $w_1=0$ or not. Fig.\ref{fig2b} shows that for bigger values
of $w_0$, $z_{cr3}$ could be smaller. At $w_0=-1$,
$z_{cr3}=z_a=3.1$.

Besides the luminosity distance, we have considered other
observables, such as the Hubble parameter $H(z)$, the linear growth
factor $D_1(z)$ and the AP parameter $H(z)d_L(z)$. The Hubble
parameter and the linear growth factor have not presented us
different degeneracy behavior with the change of the redshift in
our parameter space, however the AP parameter gave us the
interesting cosmic degeneracy which evolves with the redshift in the
interpretation of data. The property of cosmic confusion in our
parameter space brought by AP is different from that by $d_L(z)$. We
will concentrate on the AP parameter in the following.

\begin{figure}[!hbtp]
  \centering
    \subfigure[ \hspace{0.1cm} Iso-$AP$ contours with $w_1=0$ through $(\Omega_k=0,w_0=-1)$. Solid line is at $z=1.3$,
    dashed line at $z=0.2$ and dot-dashed line at $z=4$.]
    {\includegraphics[width=0.4\textwidth]{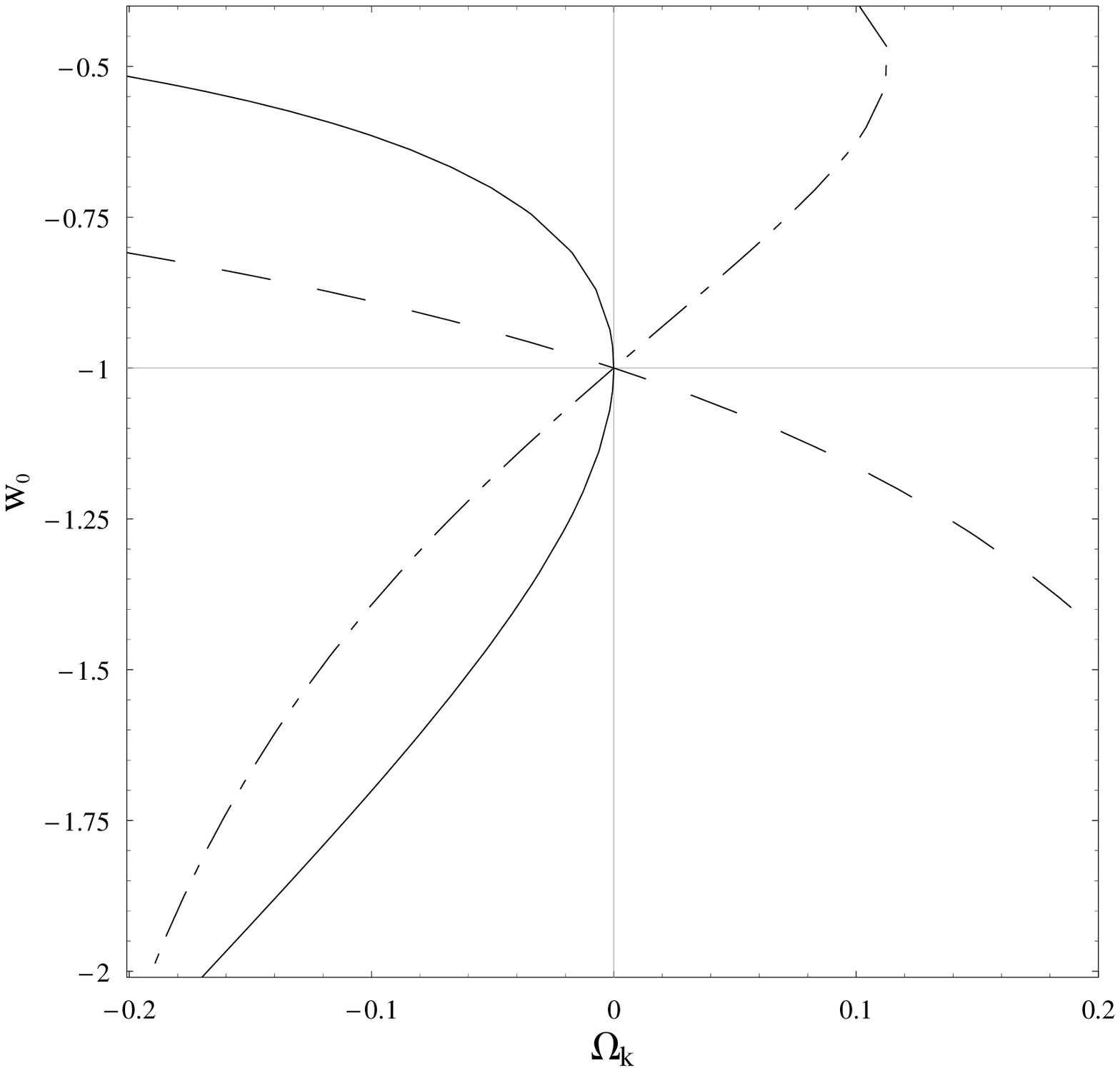}\label{fig3a}}
    \hspace{0.01\textwidth}
    \subfigure[ \hspace{0.1cm} The critical redshift to get the minimal degeneracy at $\Omega_k^0=0$ and its dependence of $w_1$.]
    {\includegraphics[width=0.4\textwidth]{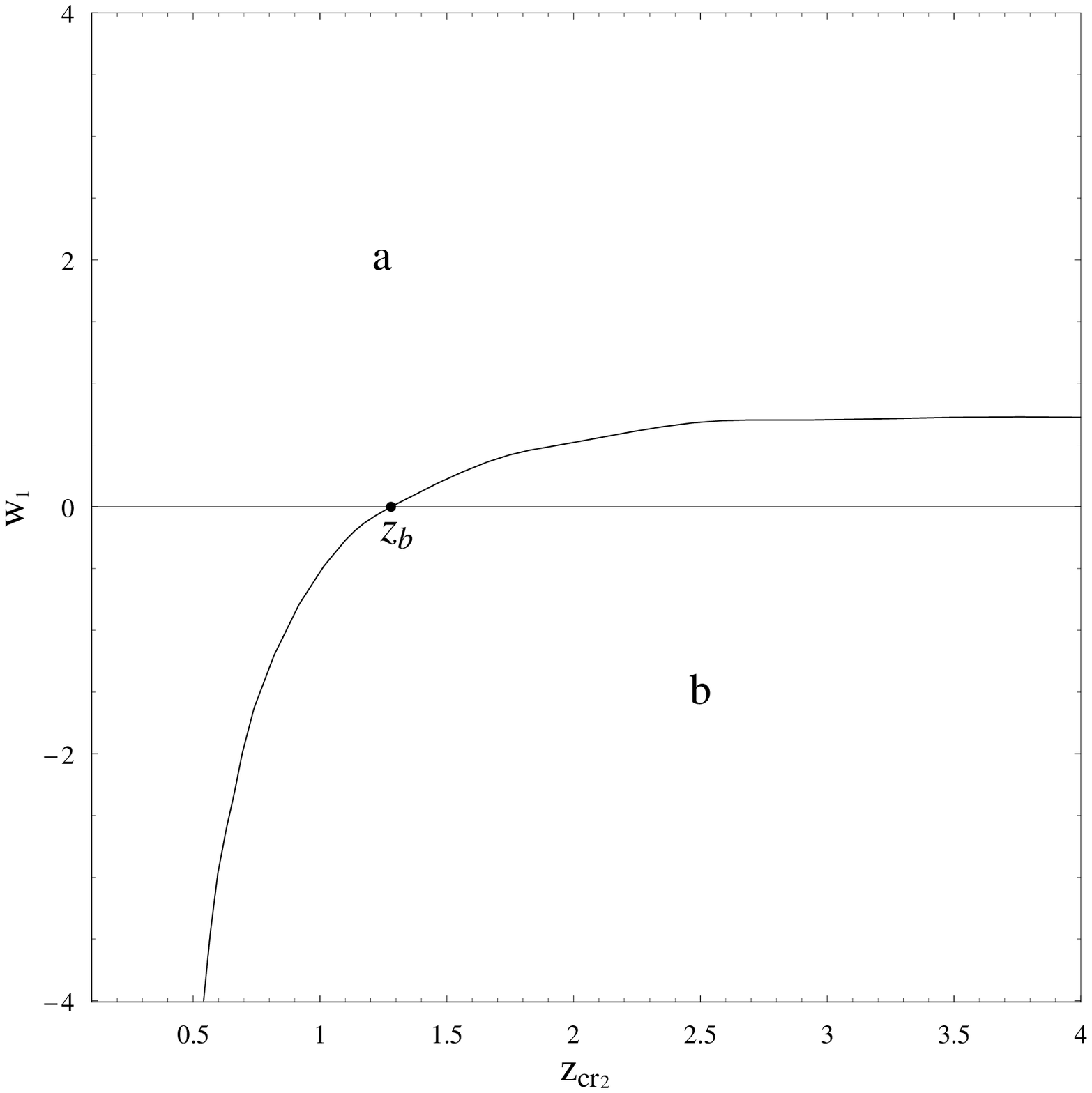}\label{fig3b}}
    \caption{.}\label{fig3}
\end{figure}

In our parameter space $(\Omega_k^0,w_0,w_1)$, for fixed $w_1=0$,
which is the constant EoS case, the iso-AP curve at low redshift
$z_1$ showed us the cosmic degeneracy in interpreting the
observation among (Closed-Q, Flat-$\Lambda$, Open-P), while at high
redshift $z_2$ the cosmic confusion arise among (Closed-P,
Flat-$\Lambda$, Open-Q). These are shown in the dashed line and
dot-dashed lines in Fig.\ref{fig3a}, respectively. There is a
crucial point at the critical redshift between $z_1$ and $z_2$,
where the degeneracy behavior changes which is shown by the solid
line in Fig.\ref{fig3a}. For $w_1=0$, this critical redshift is
found at $z_b=1.3$, where there is no degeneracy in $\Omega_k$ while
maximum degeneracy in $w_0$. This crucial point is interesting since
it can help to determine whether $\Omega_k=0$ or not.

Now let's extend the discussion to the dynamic DE described in
(\ref{para1}). We see from Fig.\ref{fig3b} that for the more
negative $w_1$ we find smaller critical redshift $z_{cr2}$ which
exhibits minimum degeneracy in $\Omega_k$ while maximum degeneracy
in $w_0$, when $w_1=0, z_{cr2}=z_b=1.3$. Regions above and below the
line in Fig.\ref{fig3b} have different degeneracy in the
interpretation of observation: degeneracy among (Closed-Q,
Flat-$\Lambda$, Open-P) in region \textbf{a} while degeneracy among
(Closed-P, Flat-$\Lambda$, Open-Q) in region \textbf{b}. They
represent the cosmic confusion measured at low and high redshifts
divided by the critical redshift for fixed $w_1$. Drastically
different degeneracies straddling the critical redshift can help to
determine whether $\Omega_k=0$. However different degeneracies in
the parameter space $(\Omega_k,w_0)$ is not the only criterion to
the spatial curvature of the universe. We will further discuss this
point in the explanation of Fig.\ref{fig7b}, \ref{fig8b} and
\ref{fig9b}.

\begin{figure}[!hbtp]
  \centering
    \subfigure[ \hspace{0.1cm} Iso-$AP$ contours with $\Omega_k=0$ through $(w_0=-1,w_1=0)$. Solid line is at $z=1.3$,
    dashed line at $z=0.2$ and dot-dashed line at $z=1.6$.]
    {\includegraphics[width=0.4\textwidth]{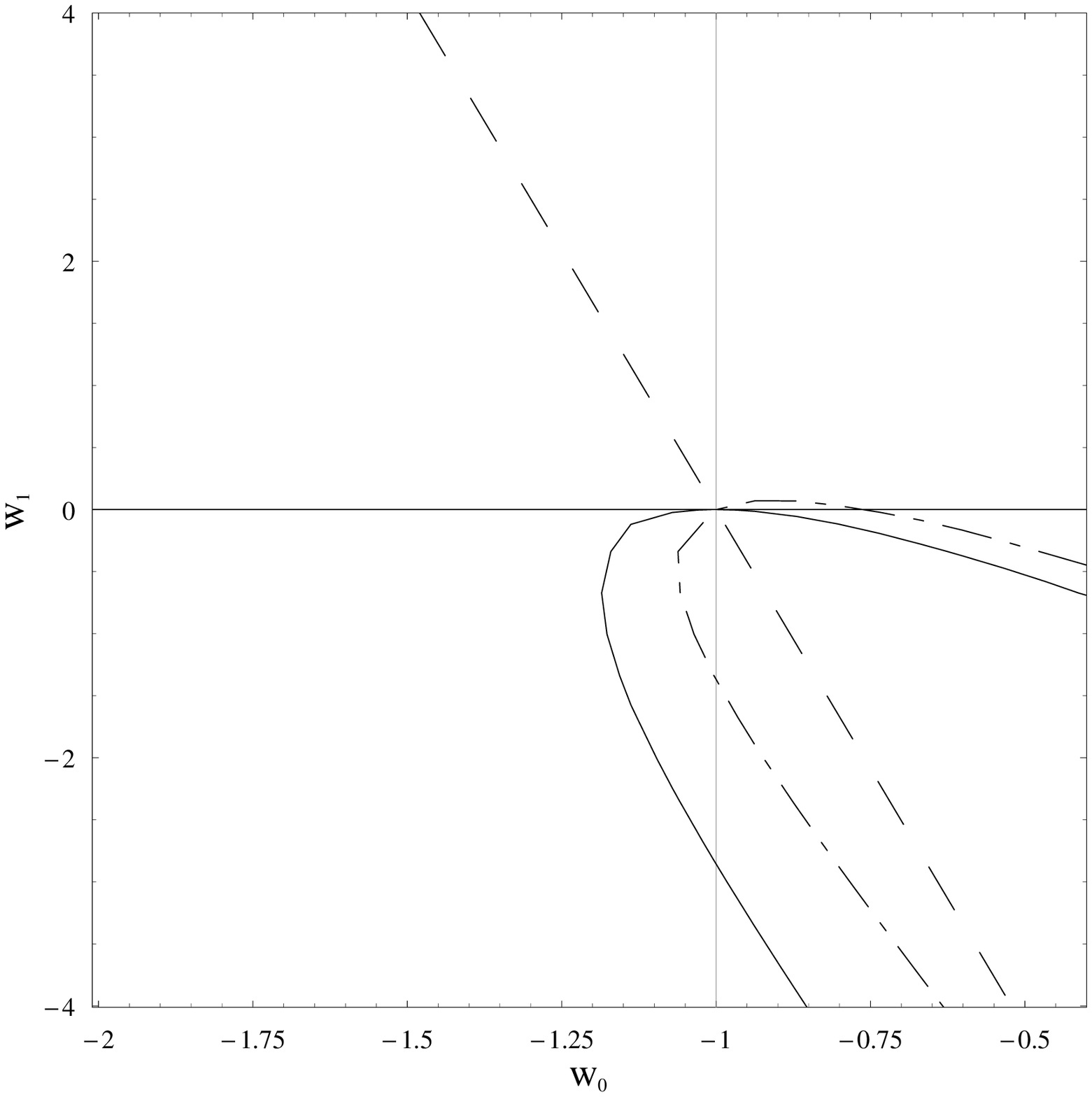}\label{fig4a}}
    \hspace{0.01\textwidth}
    \subfigure[ \hspace{0.1cm} The critical redshift to obtain minimal degeneracy at $w_1=0$ and its relation to $\Omega_k^0$.]
    {\includegraphics[width=0.4\textwidth]{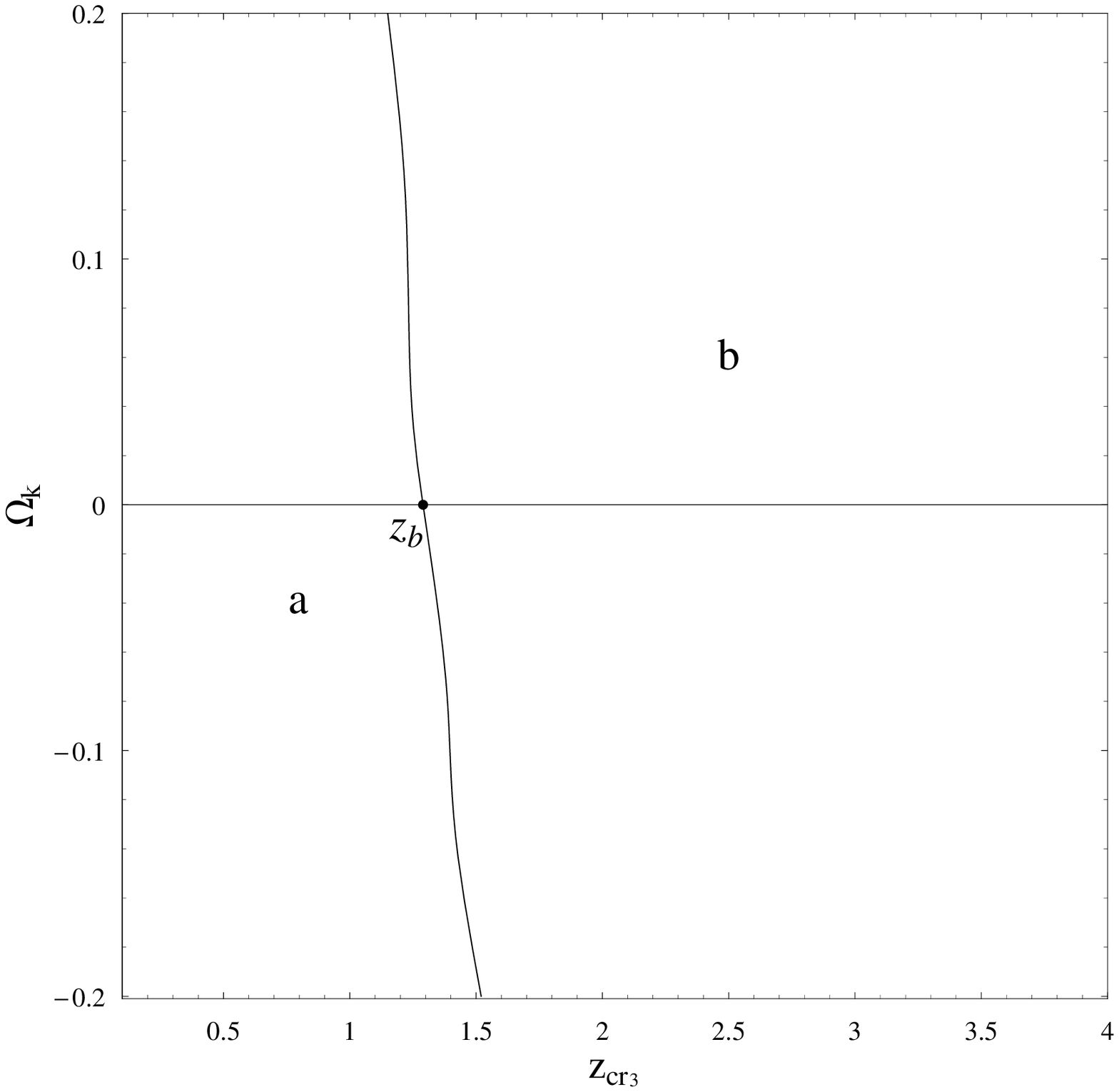}\label{fig4b}}
    \caption{.}\label{fig4}
\end{figure}

Fixing $\Omega_k$, for example assuming that our universe is flat,
from Fig.\ref{fig4a} we found that there exists a critical redshift
$z_b=1.3$ where AP parameter essentially presents us no degeneracy
in $w_1$ while maximal degeneracy with $w_0$ (the solid line). Thus
if the spatial curvature of the universe is determined, at this
critical redshift AP parameter is useful to conclude whether DE is
static or dynamic. For different chosen $\Omega_k$, this critical
redshift will change as shown in Fig.\ref{fig4b}.

\begin{figure}[!hbtp]
  \centering
    \subfigure[ \hspace{0.1cm} Iso-$AP$ contours with $\Omega_k=0$ through $(w_0=-1,w_1=0)$. Solid line is at $z=2.4$,
    dashed line at $z=1.6$ and dot-dashed line at $z=4$.]
    {\includegraphics[width=0.4\textwidth]{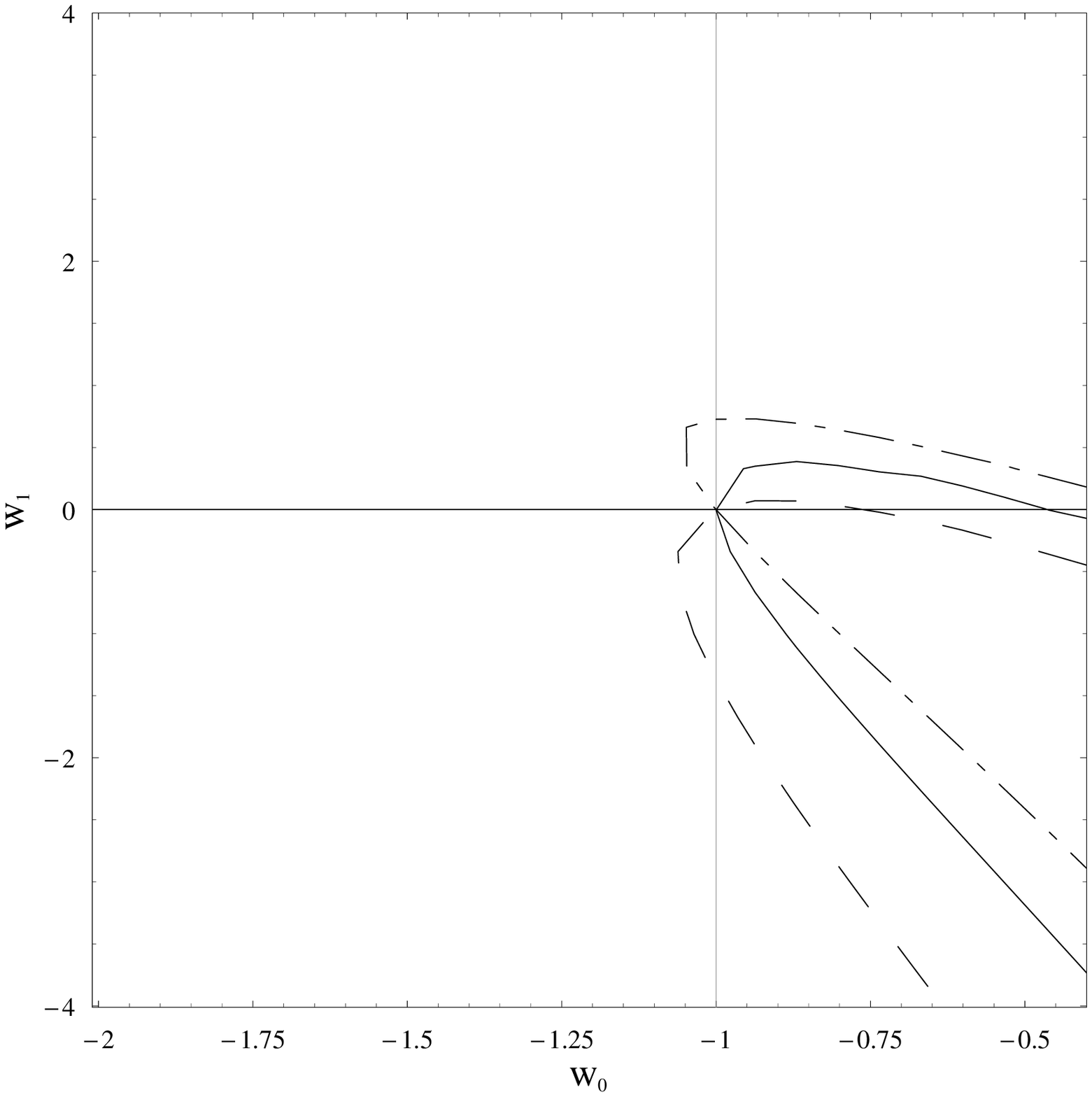}\label{fig5a}}
    \hspace{0.01\textwidth}
    \subfigure[ \hspace{0.1cm} The critical redshift to get the minimal degeneracy at $w_0=-1$ and its dependence of the spatial curvature.]
    {\includegraphics[width=0.4\textwidth]{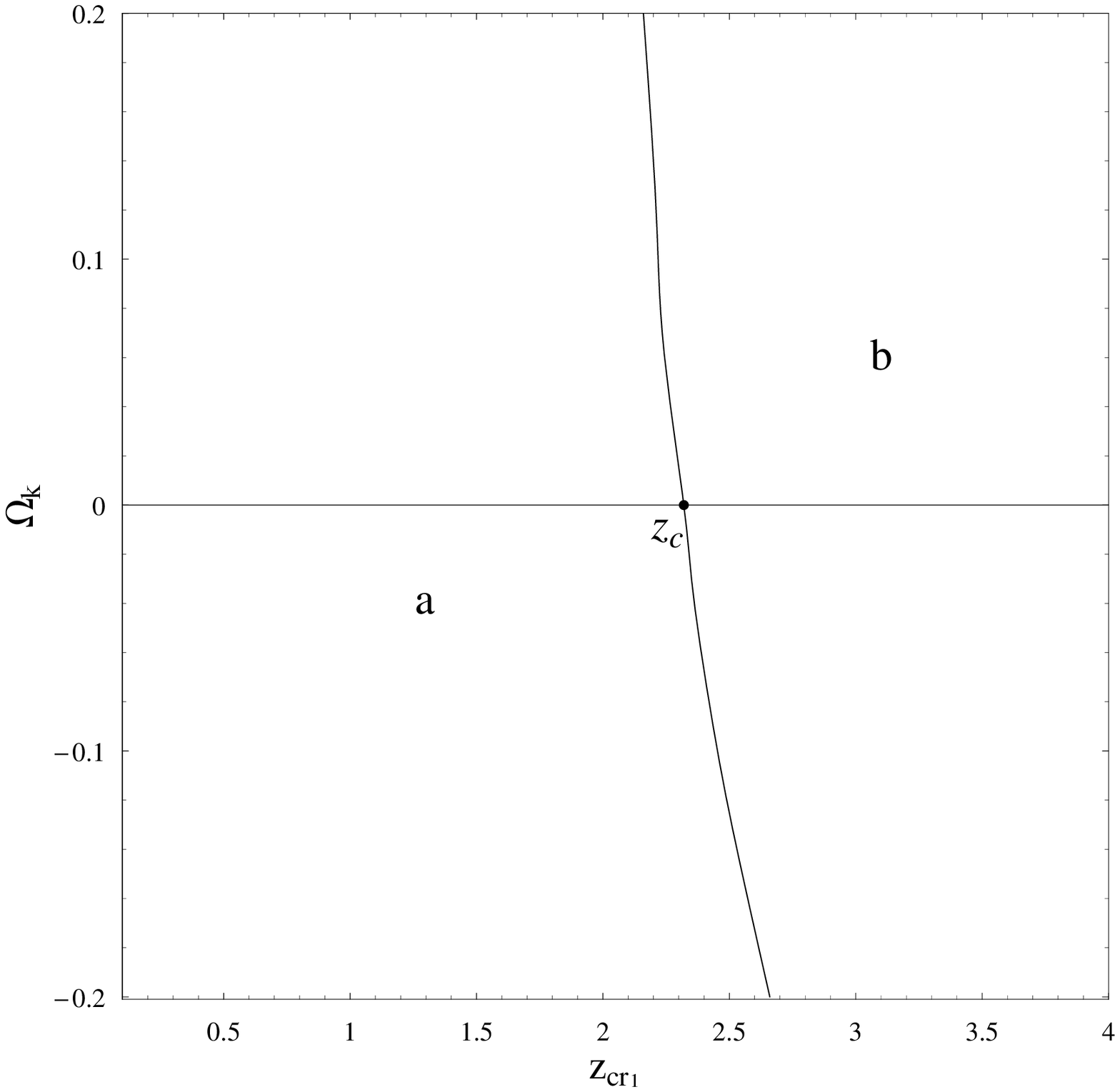}\label{fig5b}}
    \caption{.}\label{fig5}
\end{figure}

Fig.\ref{fig5a} shows that for the flat universe, we can also find a
critical redshift $z_c=2.4$ where from the AP parameter we see no
degeneracy in $w_0$, while maximum degeneracy in $w_1$ (the solid
line). Fig.\ref{fig5b} shows that in the universe with some non-zero
curvature, this critical redshift will be shifted.

\begin{figure}[!hbtp]
  \centering
    \subfigure[ \hspace{0.1cm} Iso-$AP$ contours with $w_0=-1$ through $(\Omega_k=0,w_1=0)$. Solid line is for $z=2.4$,
    dashed line for $z=0.2$ and dot-dashed line for $z=4$.]
    {\includegraphics[width=0.4\textwidth]{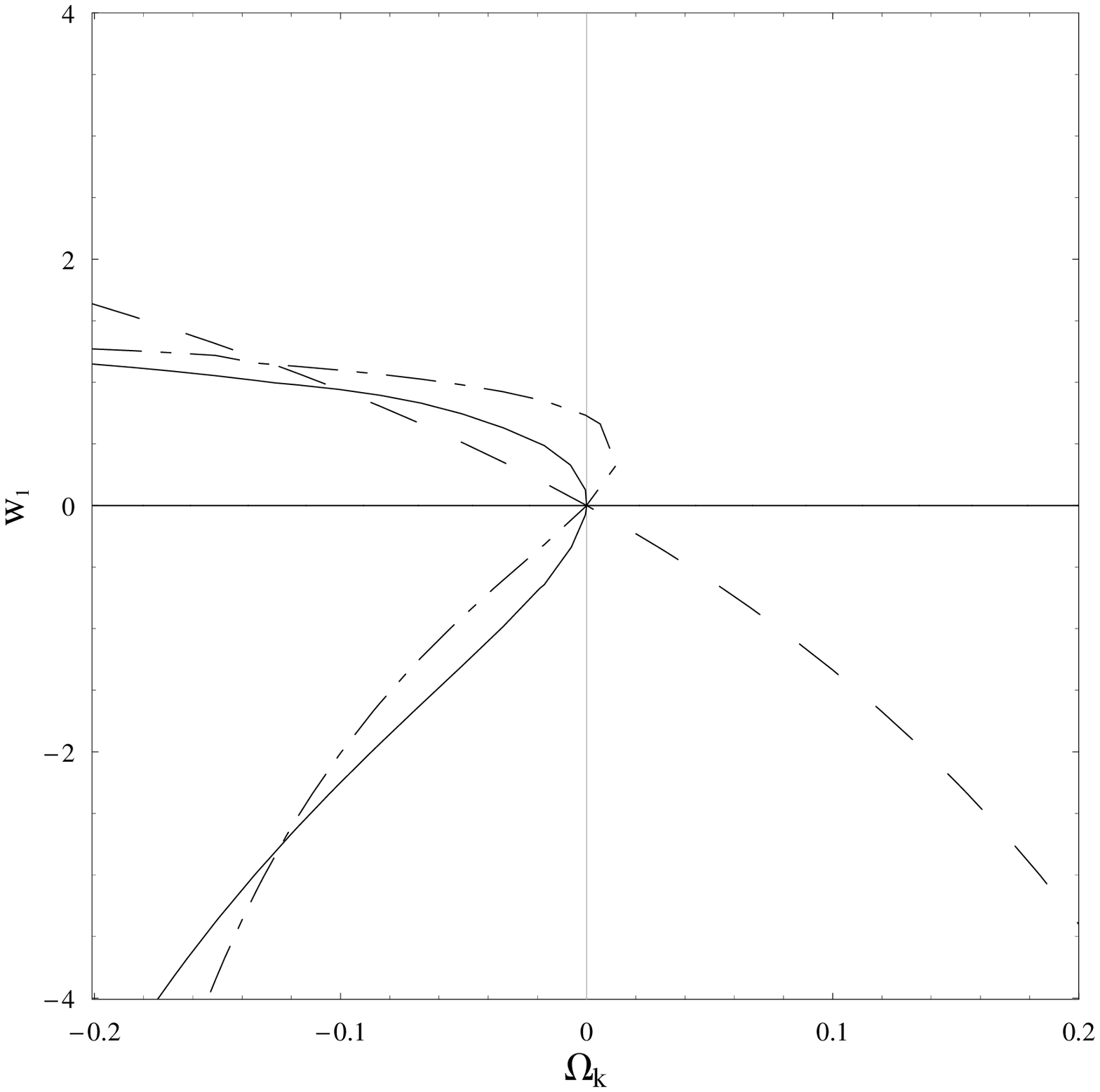}\label{fig6a}}
    \hspace{0.01\textwidth}
    \subfigure[ \hspace{0.1cm} The critical redshift to have minimal degeneracy at $\Omega_k^0=0$ and its dependence of $w_0$.]
    {\includegraphics[width=0.4\textwidth]{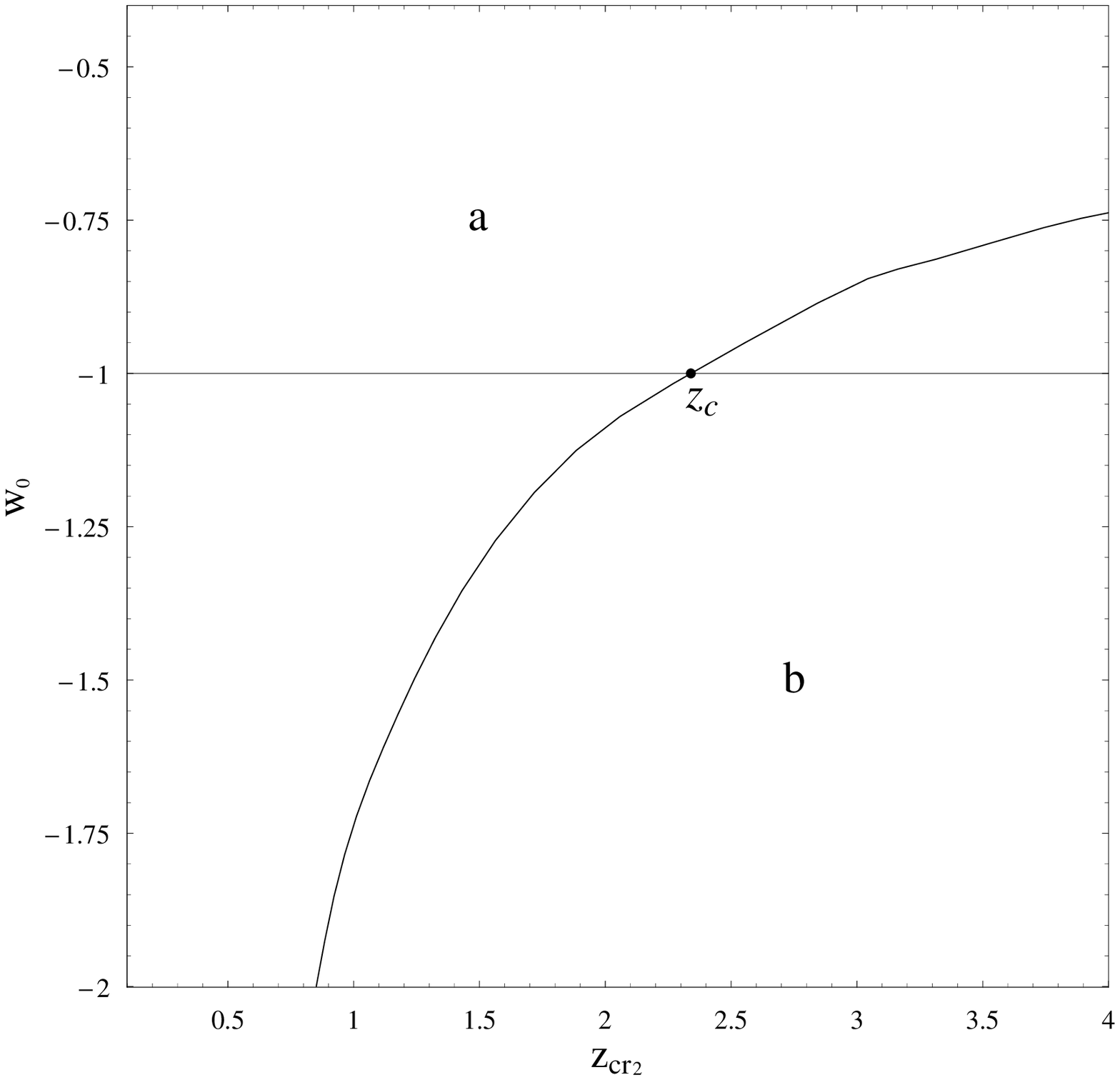}\label{fig6b}}
    \caption{.}\label{fig6}
\end{figure}

\begin{figure}[!hbtp]
  \centering
    \subfigure[ \hspace{0.1cm} Iso-$AP$ contours through $(\Omega_k=0,w_0=-1)$ at redshift $z=1$. Solid line is for $w_1=1$,
    dashed line for $w_1=-2$ and dot-dashed line for $w_1=3$.]
    {\includegraphics[width=0.4\textwidth]{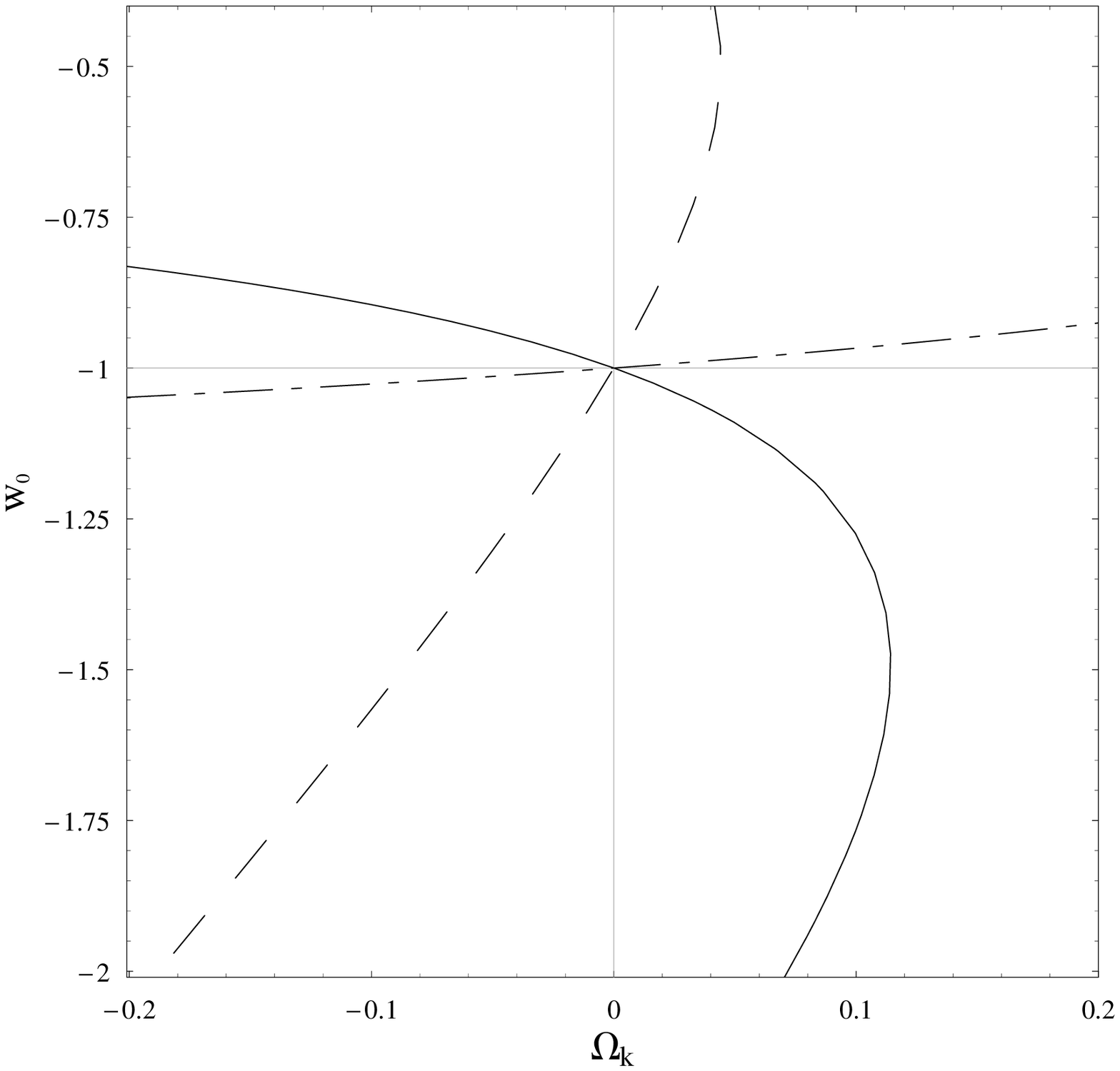}\label{fig7a}}
    \hspace{0.01\textwidth}
    \subfigure[ \hspace{0.1cm} The dashed line and solid line are for minimal degeneracy at $w_0=-1$ and $\Omega_k^0=0$, respectively.]
    {\includegraphics[width=0.4\textwidth]{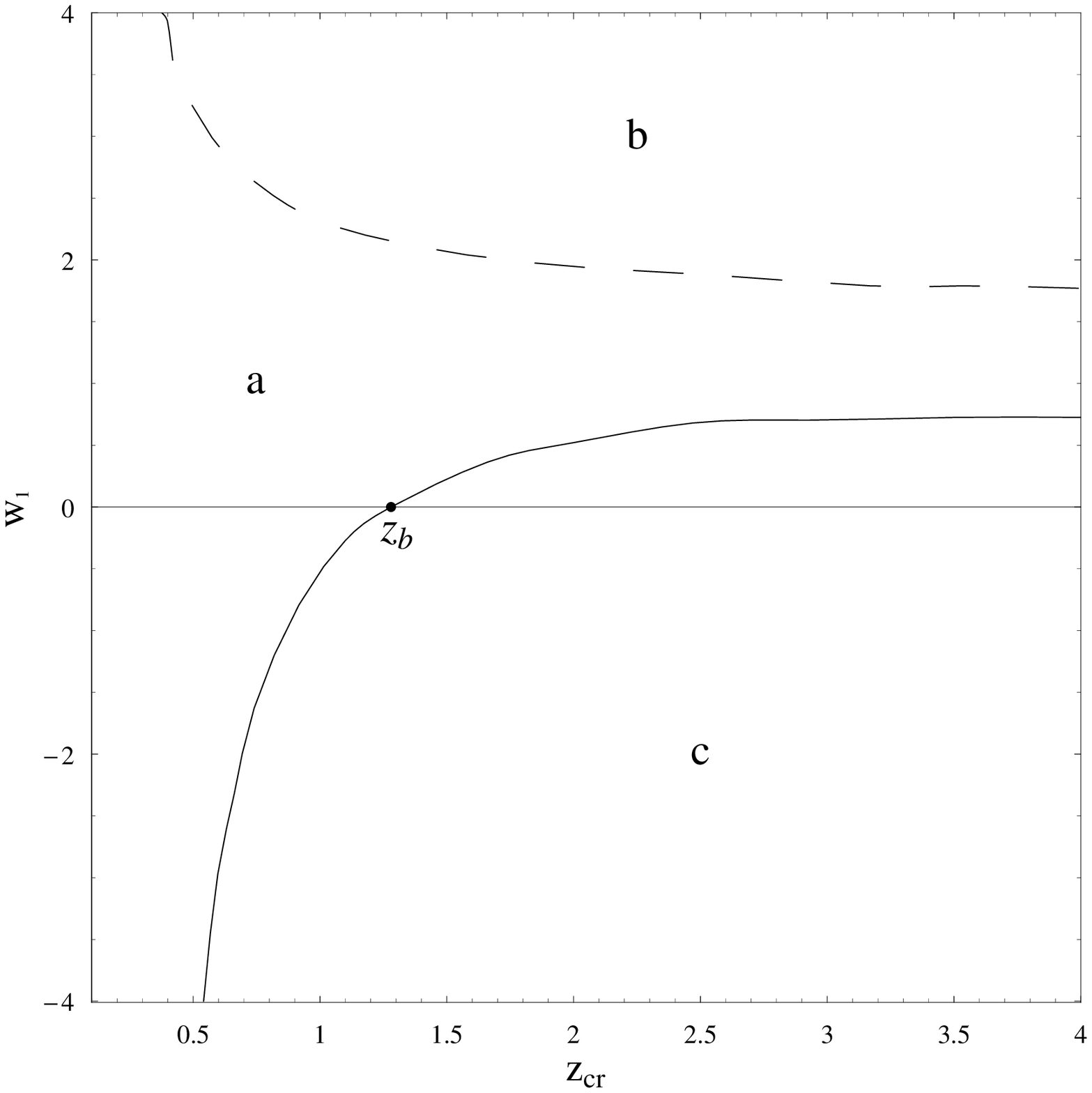}\label{fig7b}}
    \caption{.}\label{fig7}
\end{figure}

\begin{figure}[!hbtp]
  \centering
    \subfigure[ \hspace{0.1cm} Iso-$AP$ contours through $(w_0=-1,w_1=0)$ in a flat universe $\Omega_k=0$. Solid line is for  $z=1.6$,
    dashed line for $z=0.75$ and dot-dashed line for $z=3.25$.]
    {\includegraphics[width=0.4\textwidth]{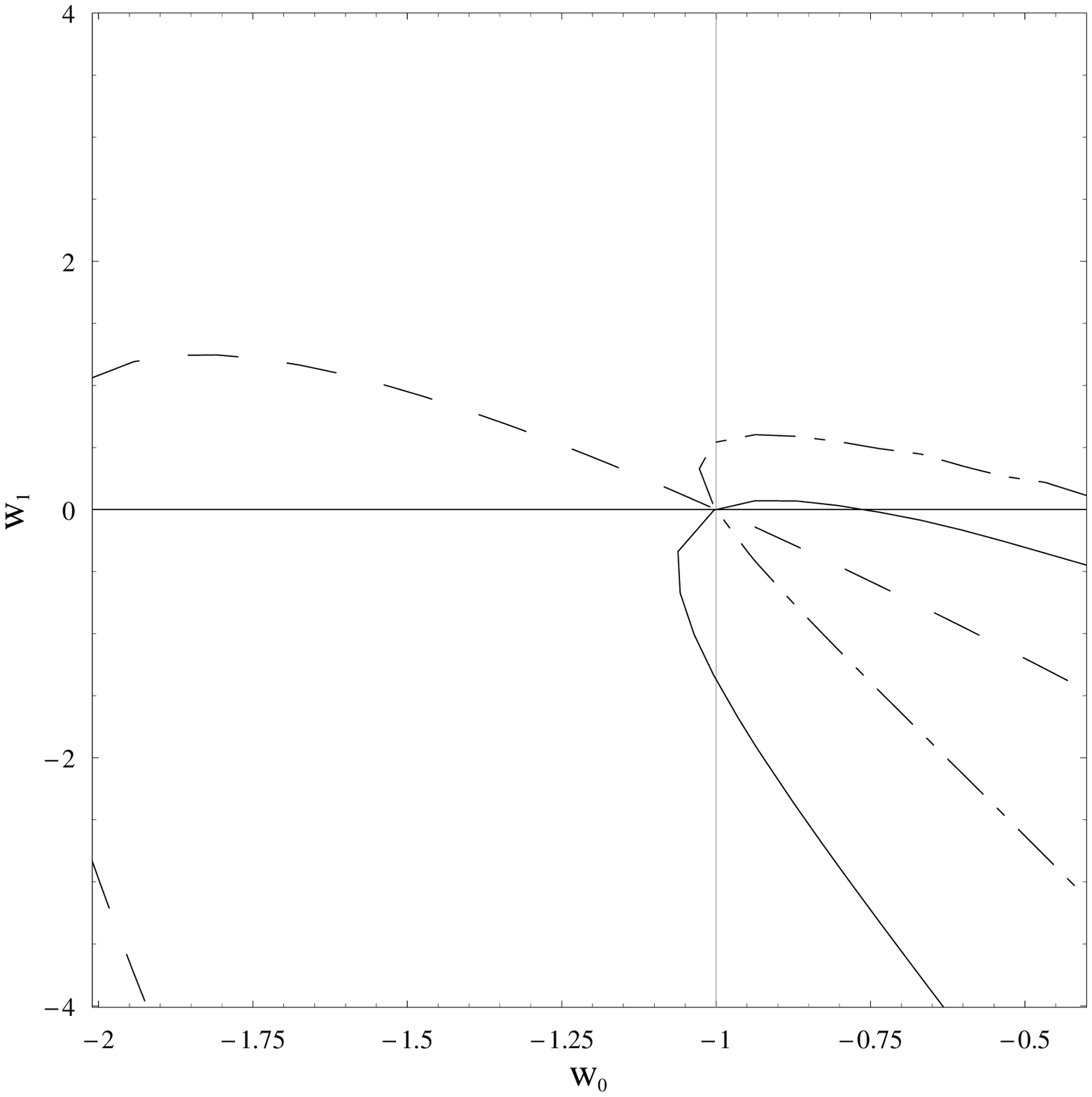}\label{fig8a}}
    \hspace{0.01\textwidth}
    \subfigure[ \hspace{0.1cm} The dashed line and solid line are for minimal degeneracy at $w_0=-1$ and $w_1=0$ respectively.]
    {\includegraphics[width=0.4\textwidth]{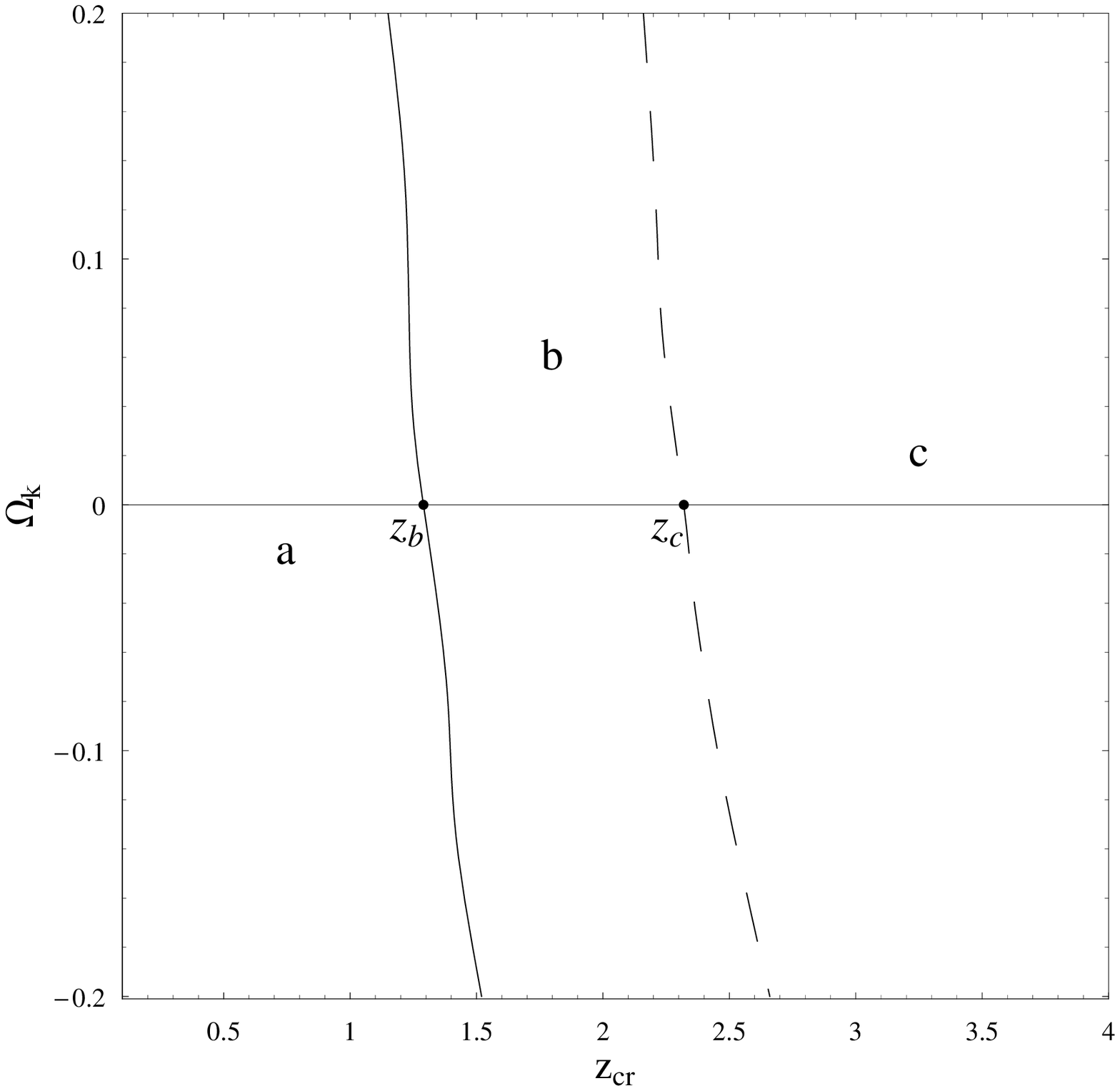}\label{fig8b}}
    \caption{.}\label{fig8}
\end{figure}

\begin{figure}[!hbtp]
  \centering
    \subfigure[ \hspace{0.1cm} Iso-$d_L$ (thin lines) and iso-$AP$ (thick lines)
    contours through $(\Omega_k=0,w_0=-1)$. Solid lines are for $z=3.25,w_1=1.25$,
    dashed lines for $z=0.65,w_1=3.25$ and dot-dashed lines for $z=2.25,w_1=-2$.]
    {\includegraphics[width=0.4\textwidth]{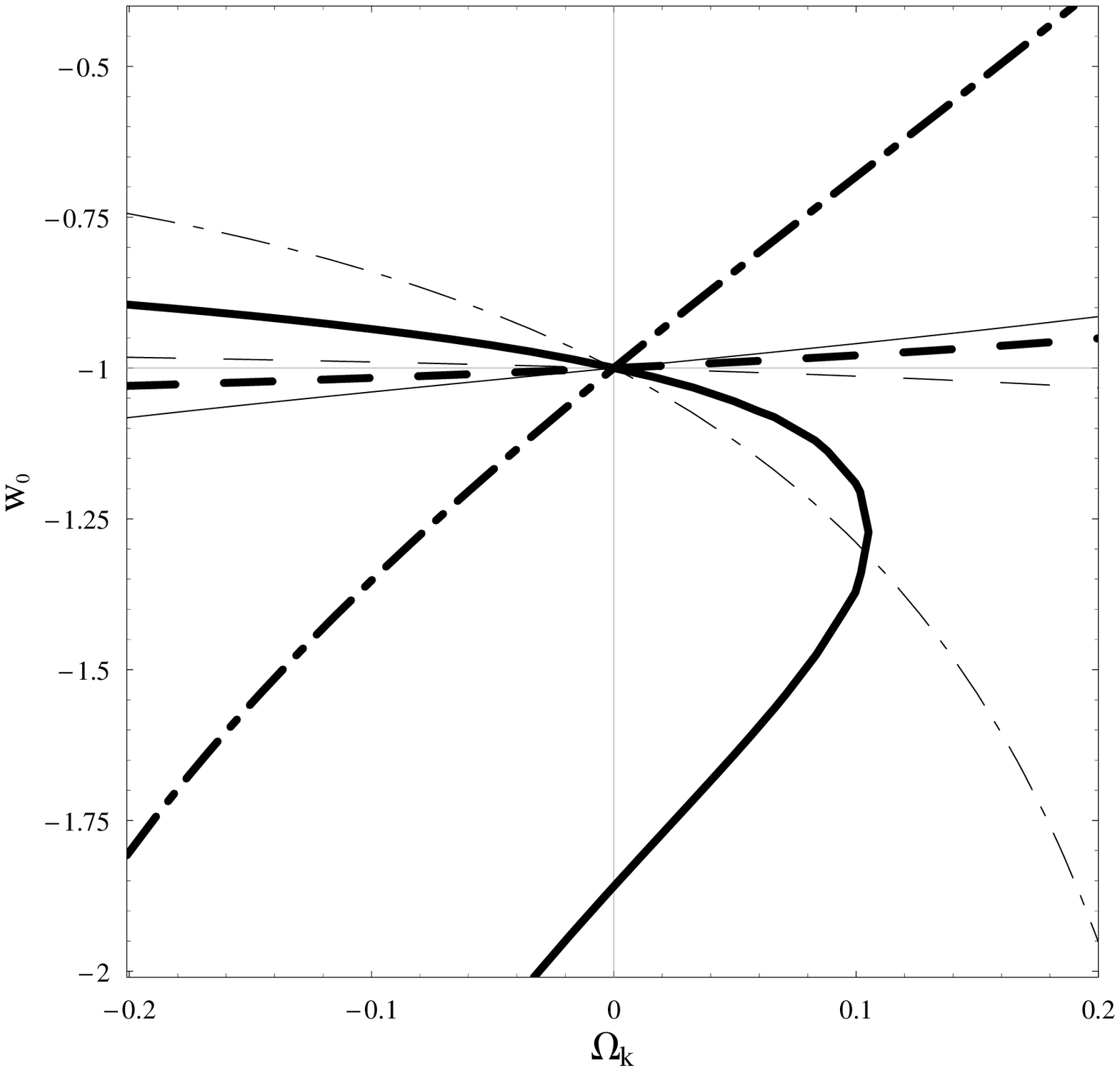}\label{fig9a}}
    \hspace{0.01\textwidth}
    \subfigure[ \hspace{0.1cm} Considering the time-dependent EoS, we have got different regions with different
cosmic degeneracies obtained from $d_L$ and $AP$. ]
    {\includegraphics[width=0.4\textwidth]{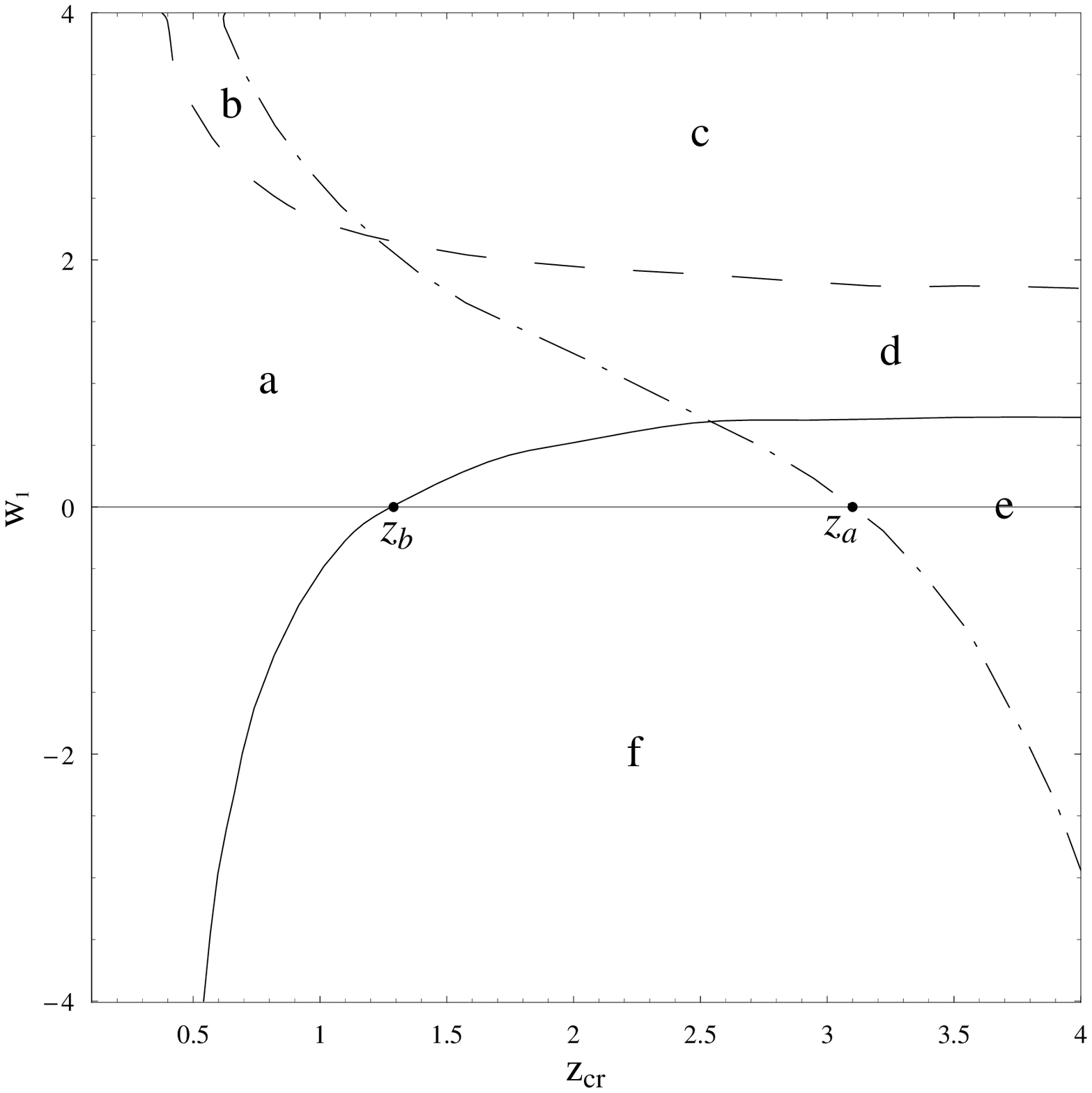}\label{fig9b}}
    \caption{.}\label{fig9}
\end{figure}

In Fig.\ref{fig6a}, we learnt that for chosen $w_0=-1$, at the
redshift $z_c=2.4$, we see no degeneracy in $\Omega_k$ (solid line).
This tells us that if we know $w_0=-1$, the AP parameter observed at
$z_c=2.4$ can help us to determine whether our universe is flat.
Arbitrary values of $w_1$ will not influence the result.
Fig.\ref{fig6b} exhibits the fact that with the change of chosen
$w_0$ values, the critical value which displays minimum degeneracy
in $\Omega_k$ in the AP parameter will change.

Until now we have discussed the AP parameter measured at different
redshifts. In Fig.\ref{fig7a} we have also exhibited the parameter
space $(\Omega_k^0,w_0)$ with the change of $w_1$ at the same
redshift $z=1$. We observed from the AP parameter measured at the
same redshift that with the increase of $w_1$ from negative value to
the small positive and big positive values, the cosmic confusions
vary from the degeneracy of (Closed-P, Flat-$\Lambda$, Open-Q)
(dashed line in Fig.\ref{fig7a}) to (CLosed-Q, Flat-$\Lambda$,
Open-P) (solid line in Fig.\ref{fig7a}) and back to (Closed-P,
Flat-$\Lambda$, Open-Q) (dot-dashed line in Fig.\ref{fig7a}).
Combining with Fig.\ref{fig3b}, we have plotted Fig.\ref{fig7b}. The
solid line in Fig.\ref{fig7b} can be used to determine whether
$\Omega_k=0$ for fixed $w_1$ as discussed in Fig.\ref{fig3b}.
Regions \textbf{a} and \textbf{c} divided by the solid line in
Fig.\ref{fig7b} have different degeneracy behaviors as explained in
Fig.\ref{fig3b}. The dashed line in Fig.\ref{fig7b} is the border
separating different cosmic degeneracy behaviors from region
\textbf{a} (Closed-Q, Flat-$\Lambda$, Open-P) to region \textbf{b}
(Closed-P, Flat-$\Lambda$, Open-Q) observed by the AP parameter with
the change of $w_1$ from small positive value to the big positive
value as described in Fig.\ref{fig7a}. From different regions in
Fig.\ref{fig7b}, which have opposite degeneracy behaviors with the
increase of $w_1$, i.e. \textbf{a}-\textbf{b} or
\textbf{a}-\textbf{c}, we can compare these opposite degeneracy at
different $w_1$ for chosen redshifts, for example comparing the
solid line with dot-dashed line or solid line with dashed line shown
in Fig.\ref{fig7a}, which could help to reduce the uncertainty in
the determination whether $\Omega_k=0$ through the observations of
AP.

Since the recent observation told us that the EoS is in the vicinity
of $w=-1$, straddling the cosmological constant boundary. In
Fig.\ref{fig8}, we have examined the cosmic degeneracy by
considering AP parameter in the vicinity of $w_0=-1$. In
Fig.\ref{fig8a} we fixed $\Omega_k=0$, and found the degeneracy in
parameter space $(w_0,w_1)$ when interpreting the AP parameter. At
low redshift, say $z=0.75$, from the dashed line in Fig.\ref{fig8a}
we obtained the degeneracy between $(w_0<-1, w_1>0)$ and
$(w_0>-1,w_1<0)$. The degeneracy behavior changes with the increase
of the redshift. When $z=1.6$, the AP parameter presents us the
degeneracy between $(w_0>-1,w_1>0)$ and $(w_0<-1,w_1<0)$ (see solid
line in Fig.\ref{fig8a}). For $z=3.25$, we have the degeneracy
between $(w_0>-1, w_1<0)$ and $(w_0<-1,w_1>0)$ (dot-dashed line in
Fig.\ref{fig8a}). For different choices of $\Omega_k$, we can also
find different degeneracies in $(w_0,w_1)$ parameter space for the
AP parameter. Combining with Fig.\ref{fig4b} and Fig.\ref{fig5b}, we
plotted our Fig.\ref{fig8b}, Different degeneracy behaviors observed
at low, intermediate and high redshifts in the parameter space
$(w_0,w_1)$ described in Fig.\ref{fig8a} are shown in regions
\textbf{a}, \textbf{b}, \textbf{c}. They are divided by the solid
line and dashed line which can be used to determine whether $w_1=0$
or $w_0=-1$ discussed in Fig.\ref{fig4b} and Fig.\ref{fig5b},
respectively. Similar to Fig.\ref{fig7b}, we can compare
observations with testing $\Omega_k$ and redshift $z$ in regions
\textbf{a} and \textbf{b} or \textbf{b} and \textbf{c} in
Fig.\ref{fig8b}, which have opposite degeneracies as solid-dashed
lines' comparison or solid-dot dashed lines' comparison as
illustrated in Fig.\ref{fig8a} , which could constrain $w_0$ and
$w_1$ around the crucial point $(w_0=-1,w_1=0)$ in Fig.\ref{fig8a}.

In the above discussion, we have concentrated on the single
observable. It is of interest to investigate sensitivities of
different observables and see whether they are complementary.
Considering the combination of $d_L$ and AP, we are going to study
how cosmic confusion can arise and be reduced. In Fig.\ref{fig9a},
we have shown the cosmic confusion that arises in the measurements
of $d_L$ and AP with thin and thick lines respectively. For
$z=3.25$, $w_1=1.25$, from the AP parameter (the thick solid line),
we learnt the degeneracy in $(\Omega_k^0,w_0)$ parameter space among
(Closed-Q, Flat-$\Lambda$, Open-P) while the degeneracy obtained
from the $d_L$ is (Closed-P, Flat-$\Lambda$, Open-Q) (see thin solid
line). For $z=2.25, w_1=-2$, the AP parameter measurement presents
us the cosmic confusion among (Closed-P, Flat-$\Lambda$, Open-Q)
(the thick dot-dashed line), however the $d_L$ observation gives the
confusion among (Closed-Q, Flat-$\Lambda$, Open-P) (thin dot-dashed
line). For $z=0.65, w_1=3.25$, the interpretations of the AP
parameter and $d_L$ also display different confusion behaviors as
(Closed-P, Flat-$\Lambda$, Open-Q) from AP (thick dashed line) and
(Closed-Q, Flat-$\Lambda$, Open-P) from $d_L$ (thin dashed line).
Different cosmic degeneracy behaviors from different observables are
possible to be used to reduce the confusion. In Fig.\ref{fig9b}, we
have combined the Fig.\ref{fig9a} with properties exhibited in
Fig.\ref{fig7b} learnt from the AP parameter and Fig.\ref{fig1b}
obtained from $d_L$. From Fig.\ref{fig7b}, we learnt that region
\textbf{a} and \textbf{f} shown in Fig.\ref{fig9b} have different
degeneracy behaviors interpreted from the AP parameter:
\textbf{a}:(Closed-Q, Flat-$\Lambda$, Open-P) and \textbf{f}:
(Closed-P, Flat-$\Lambda$, Open-Q). As exhibited in Fig.\ref{fig9a}
for $z=3.25$, $w_1=-2$, if we consider the luminosity distance, we
obtain opposite degeneracy in region \textbf{f}. Different cosmic
confusions in regions \textbf{a} and \textbf{d} have also been found
in Fig.\ref{fig1b} in the interpretation of $d_L$ and we know
\textbf{a}: (Closed-Q, Flat-$\Lambda$, Open-P) and \textbf{d}:
(Closed-P, Flat-$\Lambda$, Open-Q). If we examine the cosmic
degeneracy in region \textbf{d} from the AP, it is different from
that learnt from the observable $d_L$, which was explained in
Fig.\ref{fig9a} for $z=3.25, w_1=1.25$. The region \textbf{b}
persists the degeneracy behaviors shown in Fig.\ref{fig9a} at
$z=0.65, w_1=3.25$, which are opposite for observables AP and $d_L$.
Again, we can take advantage of the opposite degeneracies with
$w_1$'s and redshift $z$'s in region \textbf{b}, \textbf{d} or
\textbf{f} to reduce the cosmic confusion. The point that different
regions shown in Fig.\ref{fig9b} have different cosmic confusions by
considering the luminosity distance and the AP parameter is
interesting since this property shows that these two observables are
complementary and it could be helpful to reduce the uncertainty in
observation interpretation.

\begin{figure}[!hbtp]
  \centering
    \subfigure[ \hspace{0.1cm} Iso-$d_L$ contours with $1\%$ error around
    $(\Omega_k=0,w_0=-1)$. Solid lines are for $z=4,w_1=3$,
    dashed lines for $z=1.5,w_1=2$ and dot-dashed lines for $z=1,w_1=-2$.]
    {\includegraphics[width=0.4\textwidth]{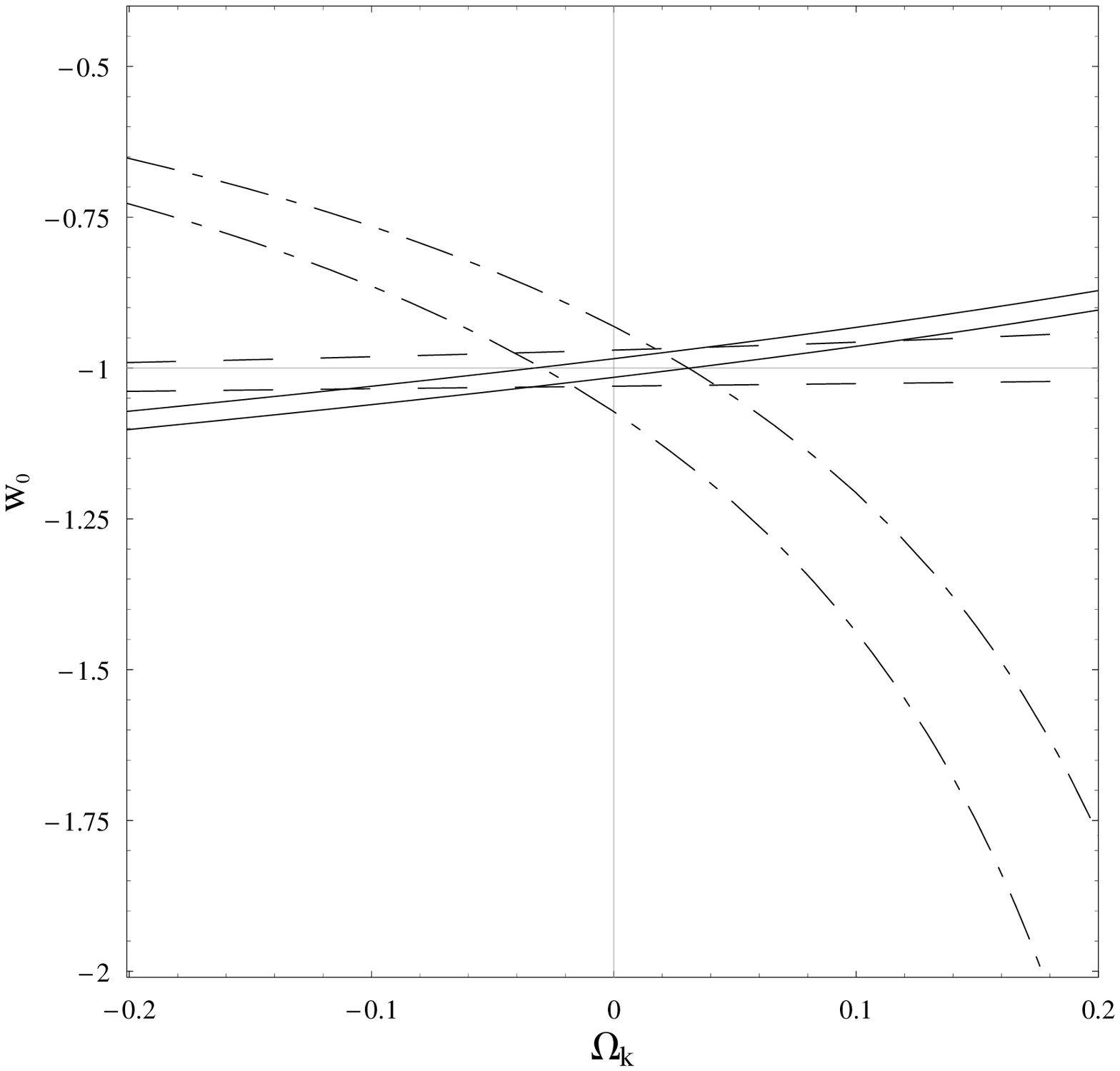}\label{fig10a}}
    \hspace{0.01\textwidth}
    \subfigure[ \hspace{0.1cm} Error sensitivity dependence of $w_1$ and the redshift.]
    {\includegraphics[width=0.4\textwidth]{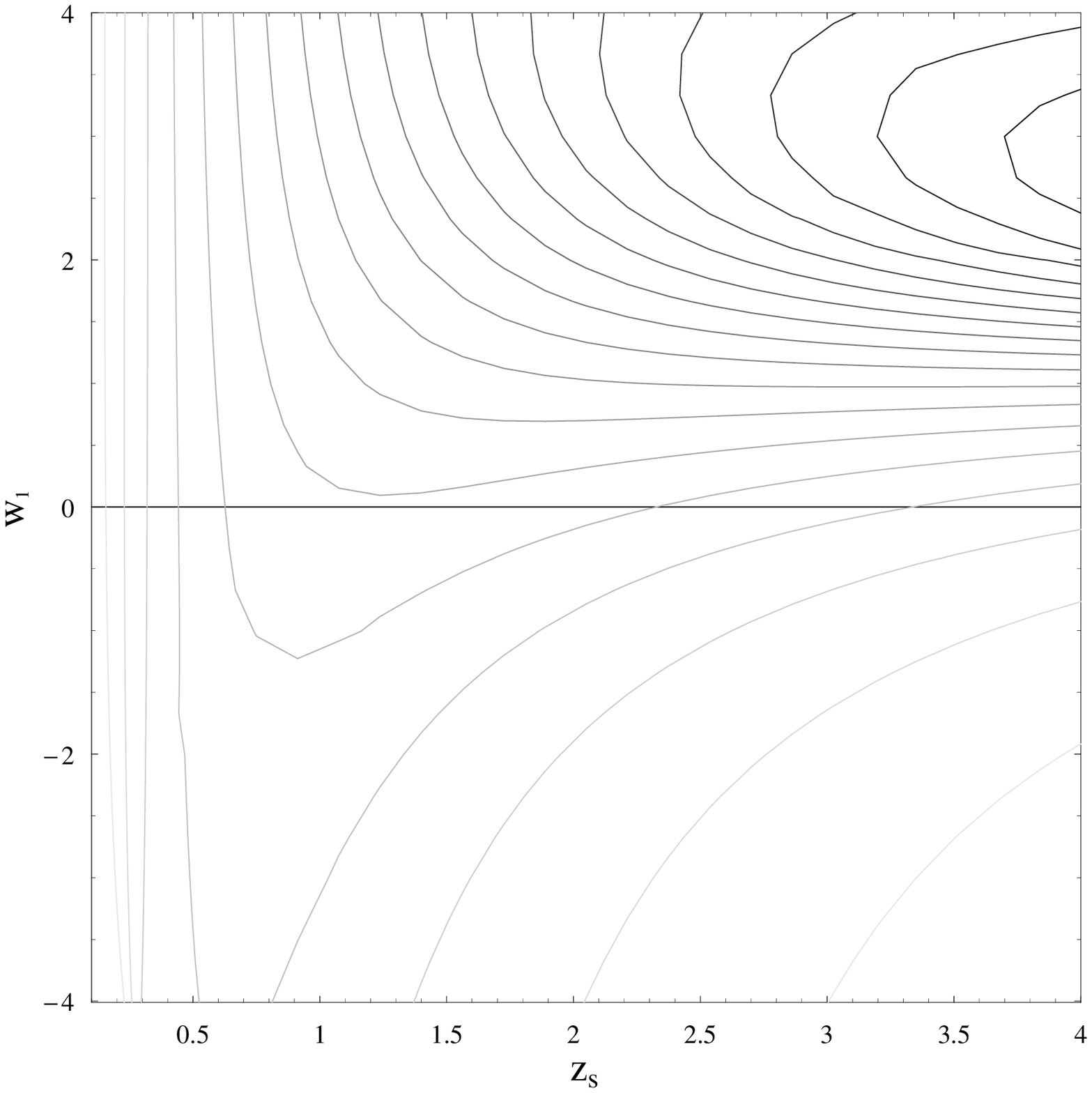}\label{fig10b}}
    \caption{.}\label{fig10}
\end{figure}

To reduce the cosmic confusion, another important factor we need to
take into account is the error  sensitivity of observables. If we
consider the relative error of an observable, for instance
$\frac{\delta d_L}{d_L}\sim 1\%$ for the luminosity distance, the
single curve of each observable will be replaced by a band. Such a
band certainly covers more than a single curve in the parameter
space and will complicate the degeneracy analysis above. The width
of the band is relevant to the redshift. At different redshift, the
same amount of error has different width in the parameter space,
which represents the error sensitivity of the observable. The
narrower the band is, the more sensitive the observable is to the
error. Mathematically the error sensitivity of $d_L$ can be
quantitatively defined as derivatives of $d_L$ with respect to
$\Omega_k$ and $w_0$ at $\Omega_k=0$, $w_0=-1$,
\begin{equation}\label{error}
    \xi_{d_L}(z,w_1)=\frac{\partial^2 d_L(z,\Omega_k,w_0,w_1)}{\partial\Omega_k\partial
    w_0}|_{\Omega_k=0,w_0=-1},
\end{equation}
and the large value of the $\xi$ corresponds to high sensitivity.
The values of $\xi_{d_L}(z,w_1)$ result in the choices in the
parameter space $(z,w_1)$. As shown in Fig.\ref{fig10b}, sensitivity
decreases with the brightness of contour lines, thus the error
sensitivity is high at high $z$ for $w_1>0$. In Fig.\ref{fig10a},
sample plot shows that with the same relative error, the higher the
sensitivity, the narrower the error band in $(\Omega_k^0,w_0)$
parameter space for $d_L$.

\begin{figure}[!hbtp]
  \centering
    \subfigure[ \hspace{0.1cm} Iso-$d_L$ contours with $1\%$ error around
    $(w_0=-1,w_1=0)$. Solid lines are for $z=1,\Omega_k=-0.2$,
    dashed lines for $z=1,\Omega_k=0$ and dot-dashed lines for $z=3,\Omega_k=0.15$.]
    {\includegraphics[width=0.4\textwidth]{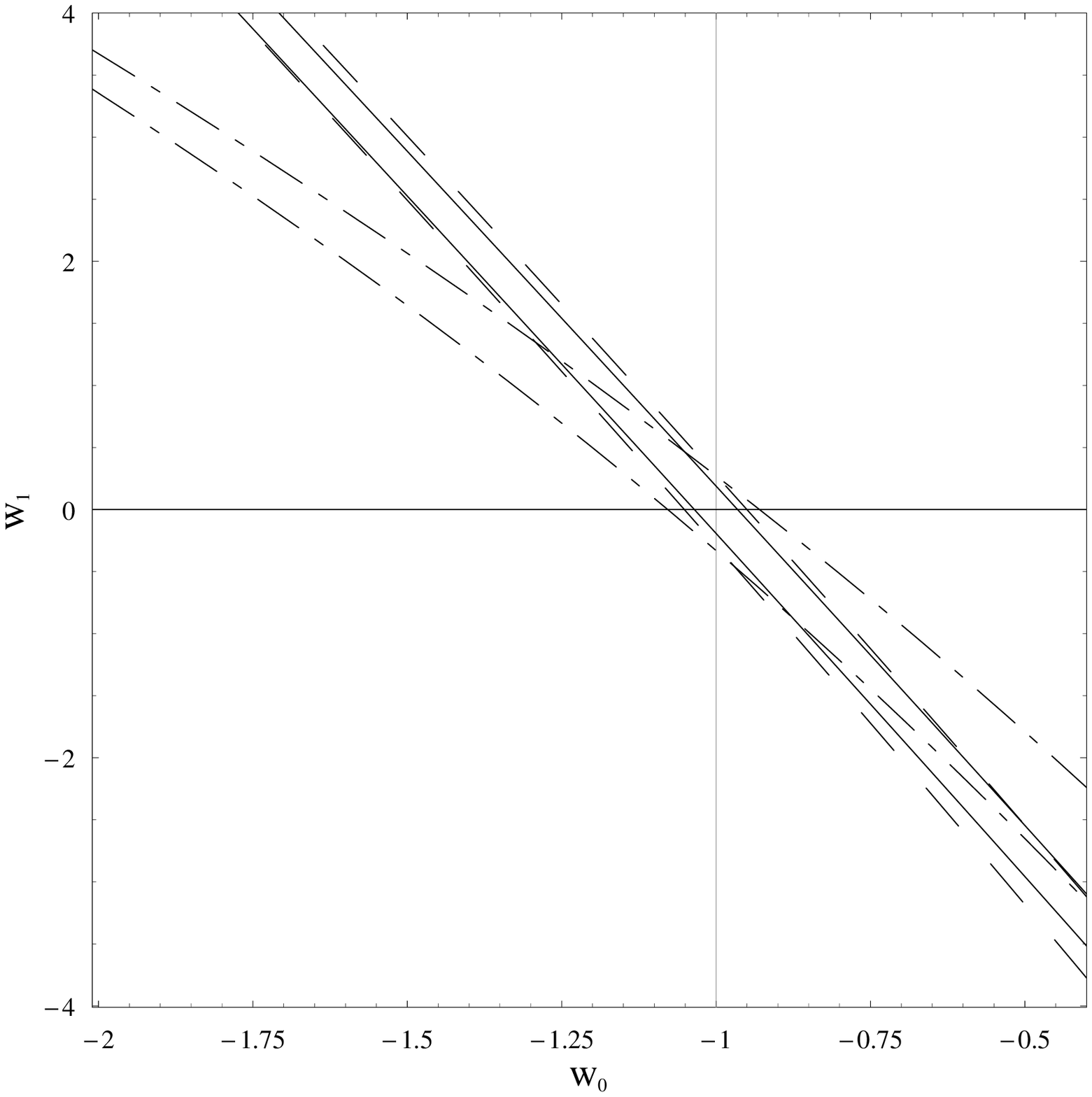}\label{fig11a}}
    \hspace{0.01\textwidth}
    \subfigure[ \hspace{0.1cm} Error sensitivity dependence of $\Omega_k^0$ and the redshift.]
    {\includegraphics[width=0.4\textwidth]{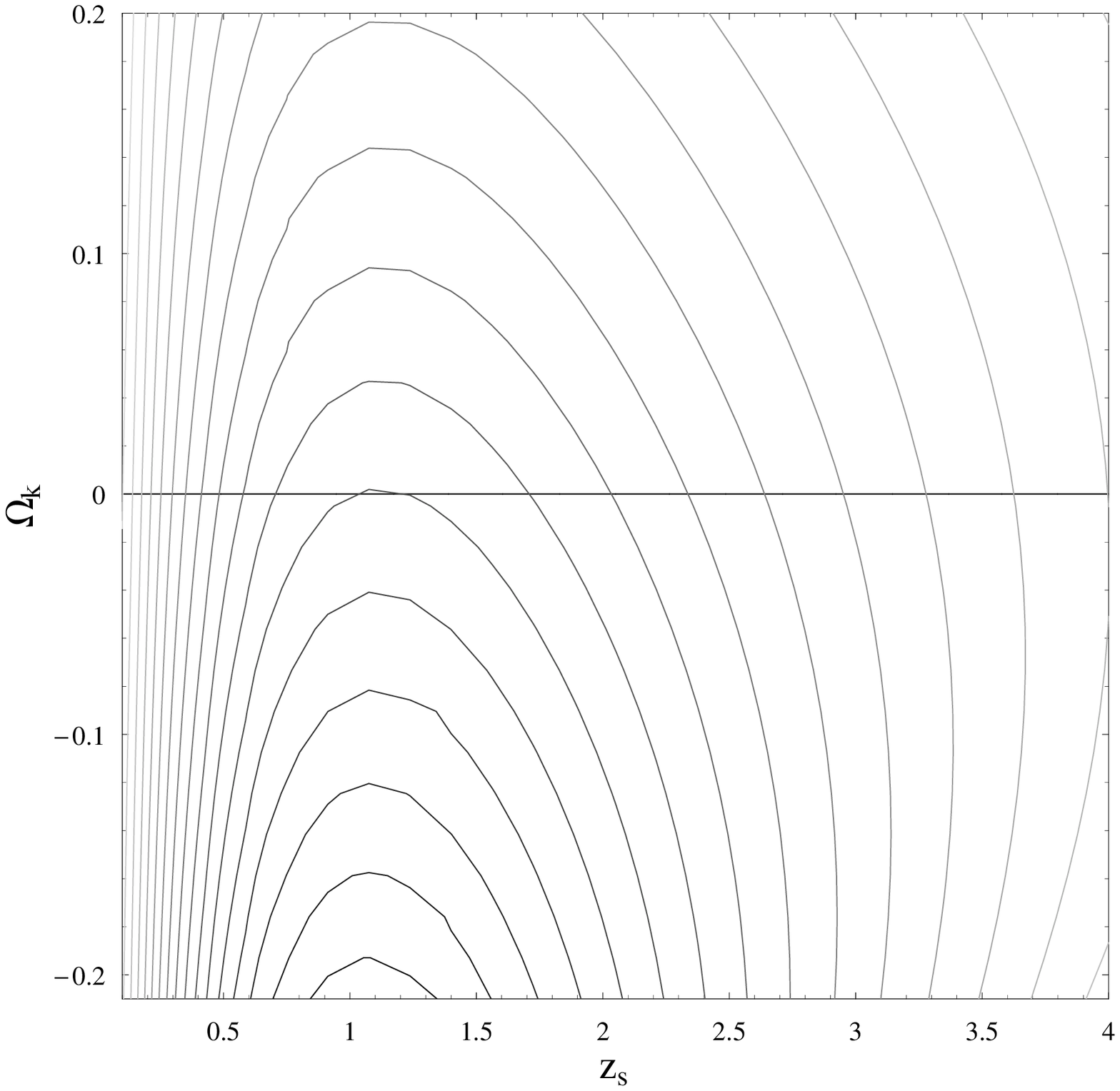}\label{fig11b}}
    \caption{.}\label{fig11}
\end{figure}

\begin{figure}[!hbtp]
  \centering
    \subfigure[ \hspace{0.1cm} Iso-$d_L$ contours with $1\%$ error around
    $(\Omega_k=0,w_1=0)$. Solid lines are for $z=4,w_0=-0.4$,
    dashed lines for $z=1.5,w_0=-0.75$ and dot-dashed lines for $z=0.5,w_0=-2$.]
    {\includegraphics[width=0.4\textwidth]{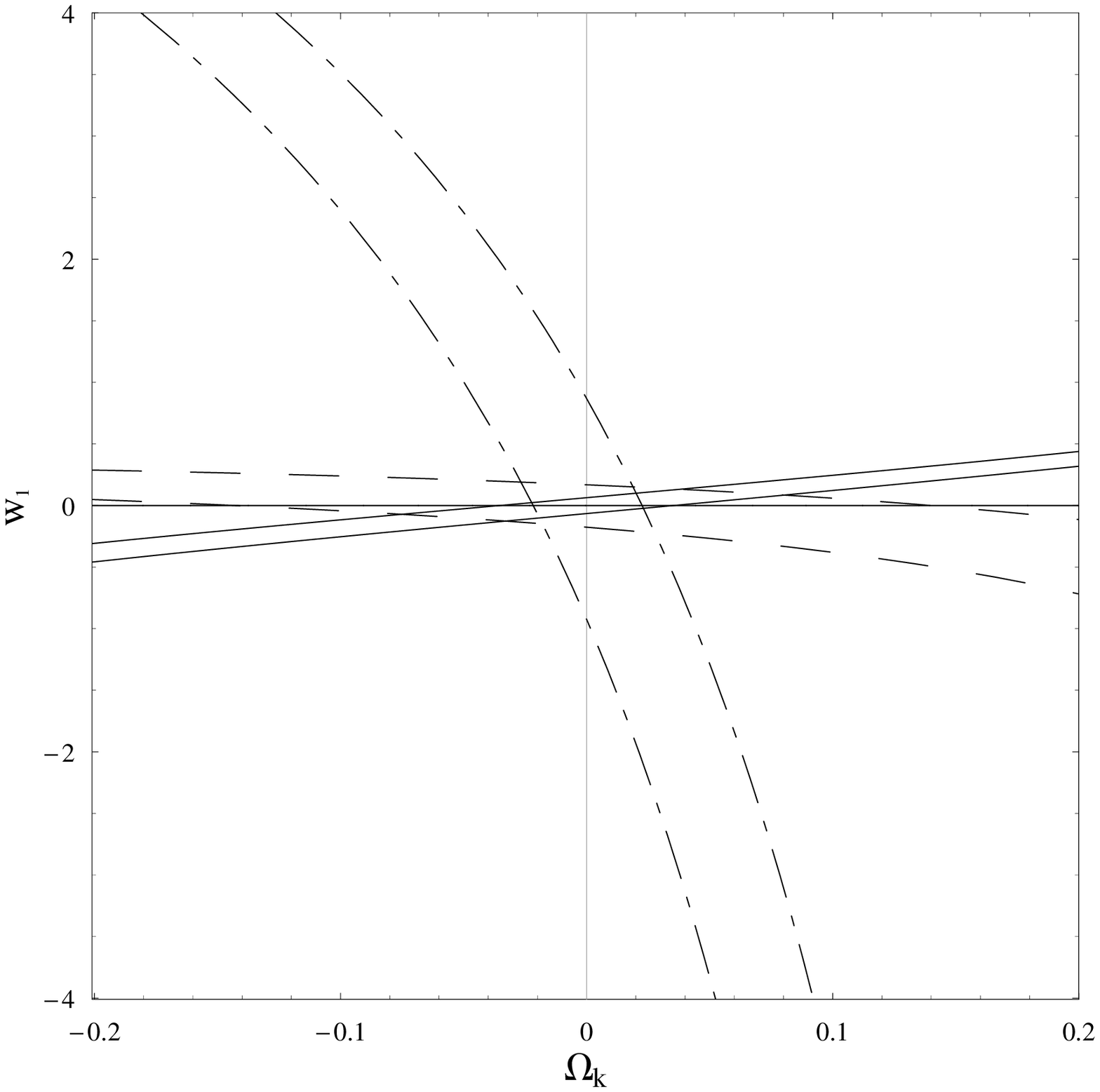}\label{fig12a}}
    \hspace{0.01\textwidth}
    \subfigure[ \hspace{0.1cm} Error sensitivity depnedence of $w_0$ and the redshift.]
    {\includegraphics[width=0.4\textwidth]{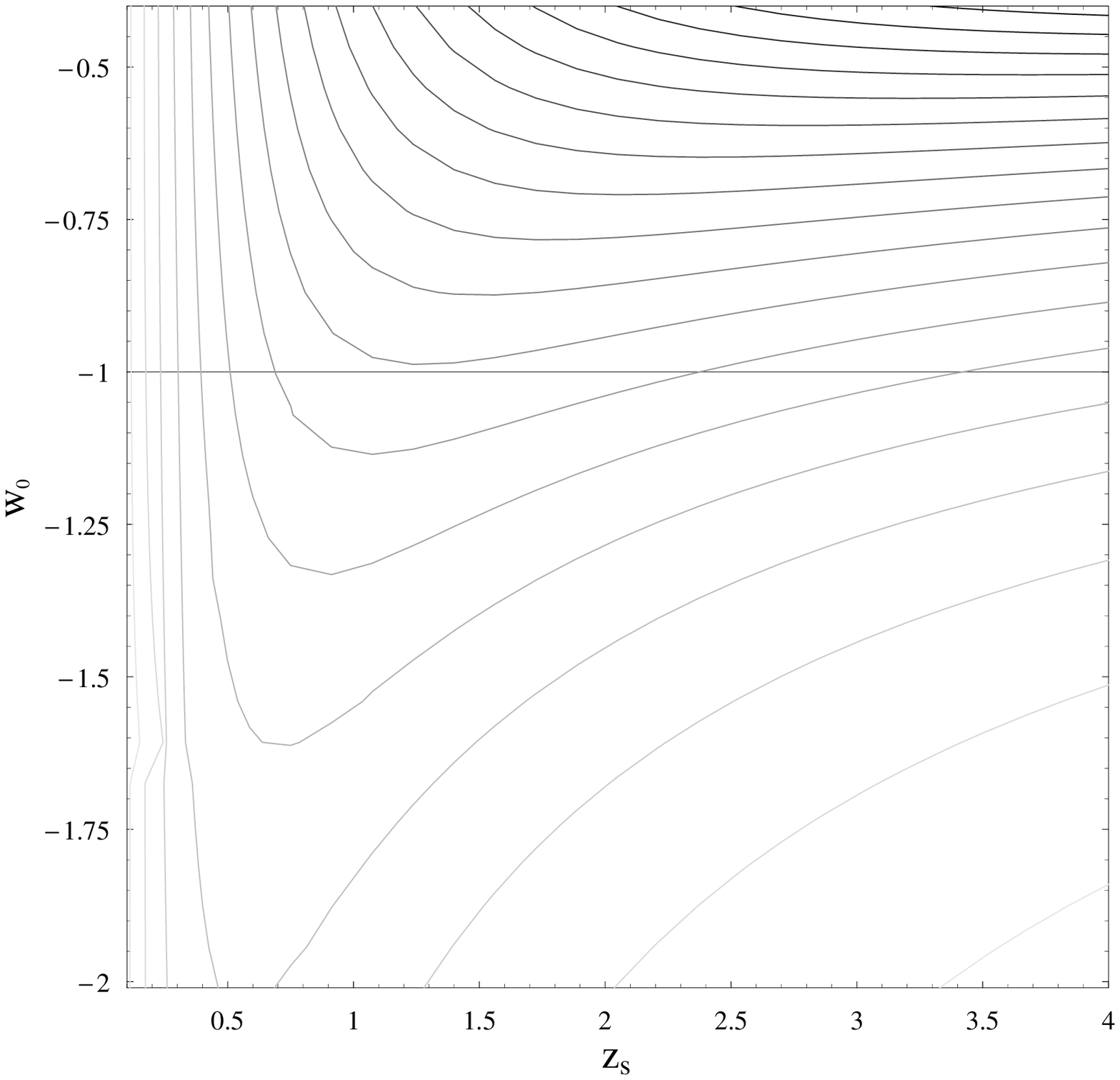}\label{fig12b}}
    \caption{.}\label{fig12}
\end{figure}

Fig.\ref{fig11} shows that for the same relative error, the band in
the parameter space $(w_0,w_1)$ for $d_L$ is narrower with positive
curvature and small redshift. Similarly Fig.\ref{fig12} displays the
narrower band in the parameter space $(\Omega_k,w_1)$ for $d_L$ at
high $z$ and $w_0>-1$.

\begin{figure}[!hbtp]
  \centering
    \subfigure[ \hspace{0.1cm} Iso-$AP$ contours with $1\%$ error around
    $(\Omega_k=0,w_0=-1)$. Solid lines are for $z=4,w_1=4$,
    dashed lines for $z=1,w_1=4$ and dot-dashed lines for $z=0.5,w_1=-4$.]
    {\includegraphics[width=0.4\textwidth]{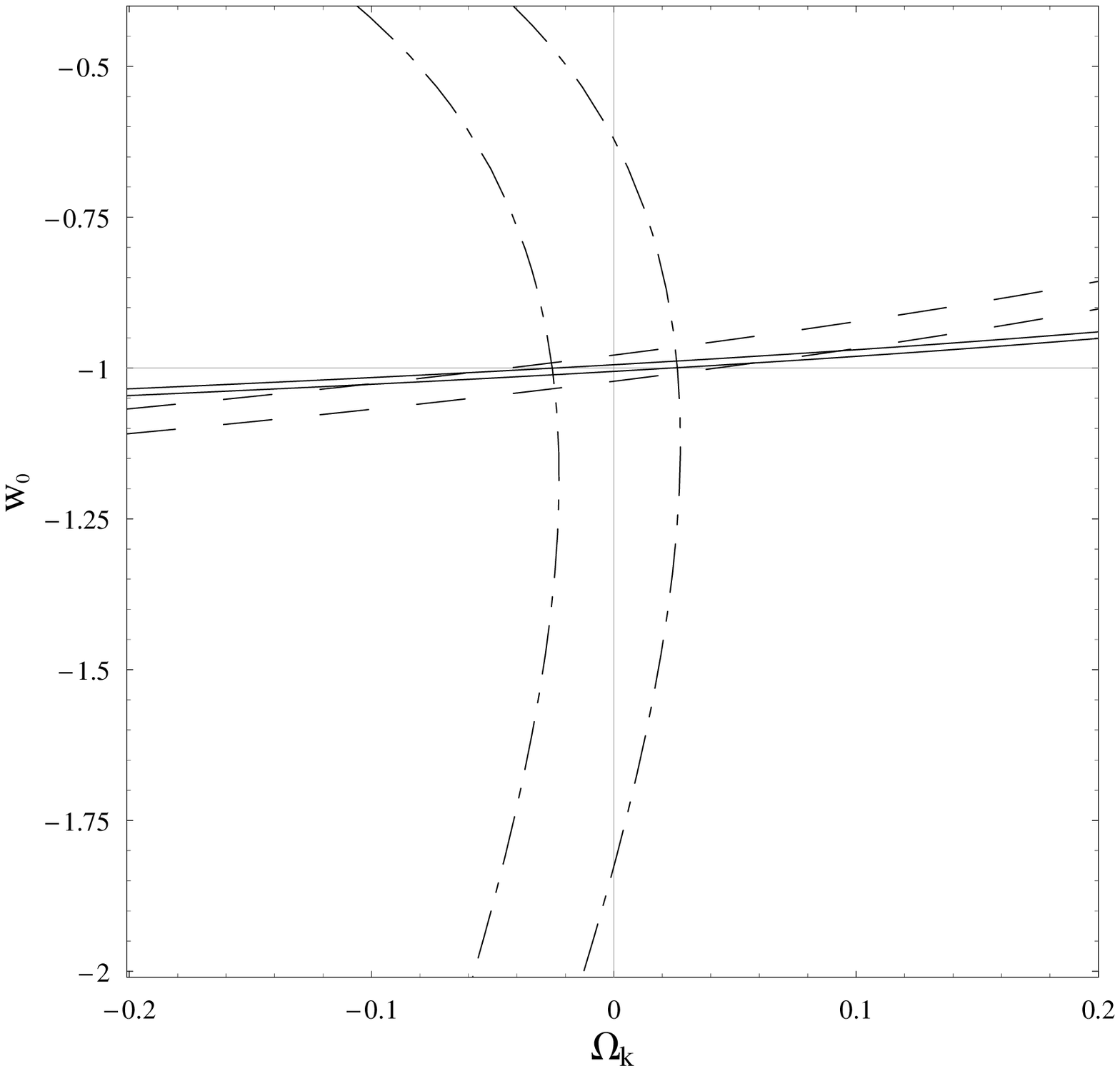}\label{fig13a}}
    \hspace{0.01\textwidth}
    \subfigure[ \hspace{0.1cm} Error sensitivity dependence of $w_1$ and the redshift.]
    {\includegraphics[width=0.4\textwidth]{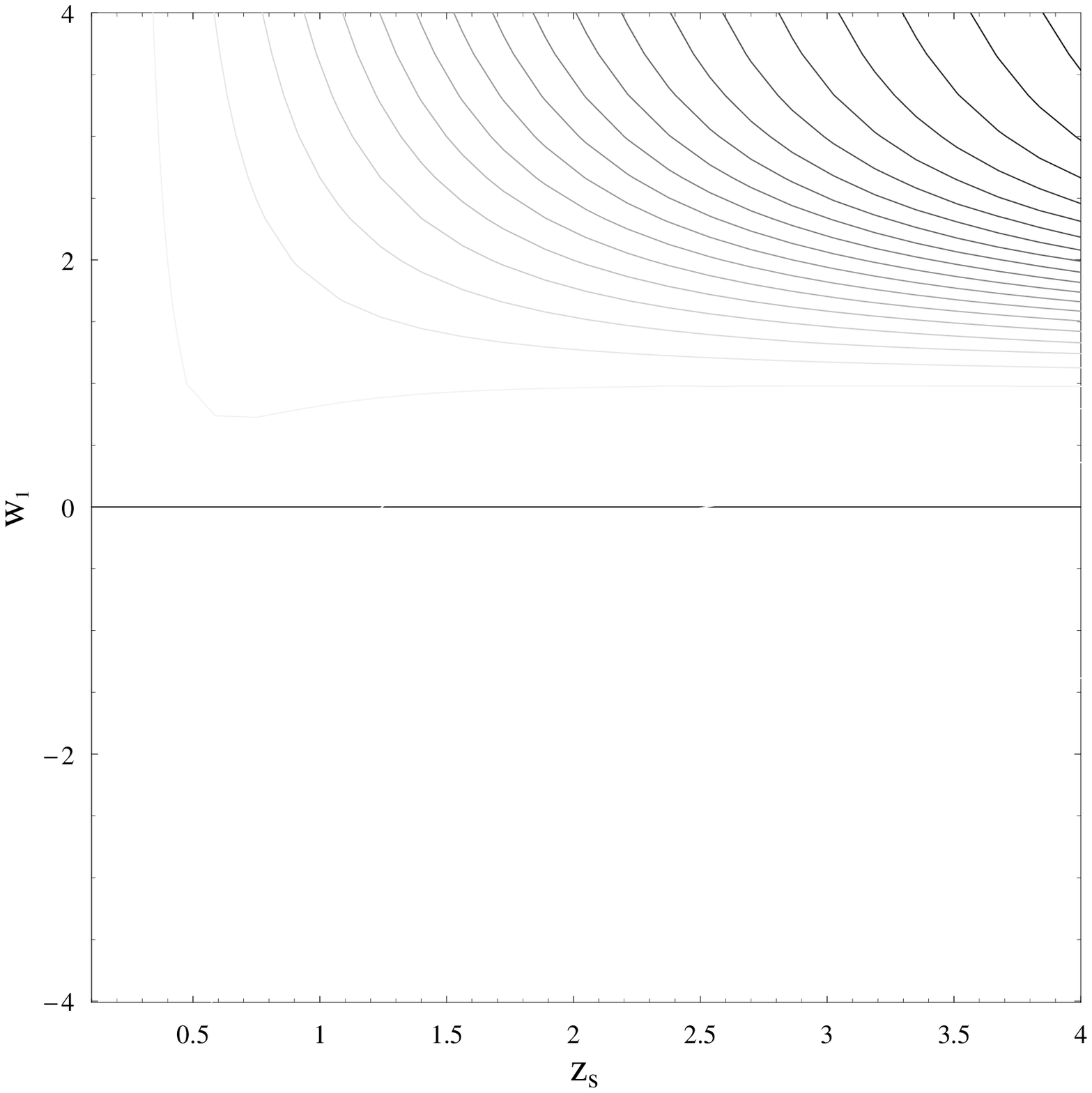}\label{fig13b}}
    \caption{.}\label{fig13}
\end{figure}

\begin{figure}[!hbtp]
  \centering
    \subfigure[ \hspace{0.1cm} Iso-$AP$ contours with $1\%$ error around
    $(w_0=-1,w_1=0)$. Solid lines are for $z=0.75,\Omega_k=-0.2$,
    dashed lines for $z=0.75,\Omega_k=0$ and dot-dashed lines for $z=3.5,\Omega_k=0.1$.]
    {\includegraphics[width=0.4\textwidth]{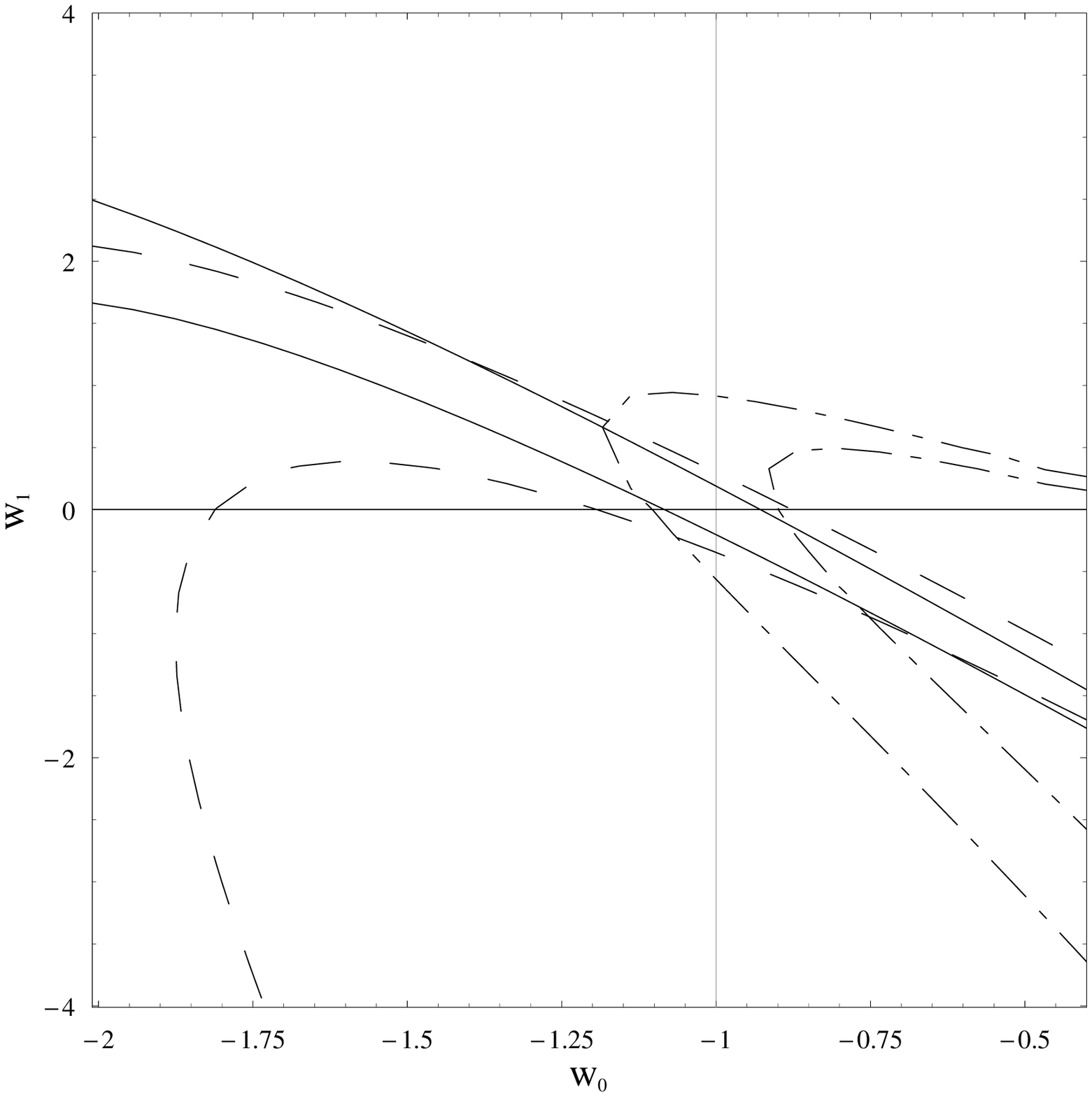}\label{fig14a}}
    \hspace{0.01\textwidth}
    \subfigure[ \hspace{0.1cm} Error sensitivity dependence on $\Omega_k^0$ and the redshift.]
    {\includegraphics[width=0.4\textwidth]{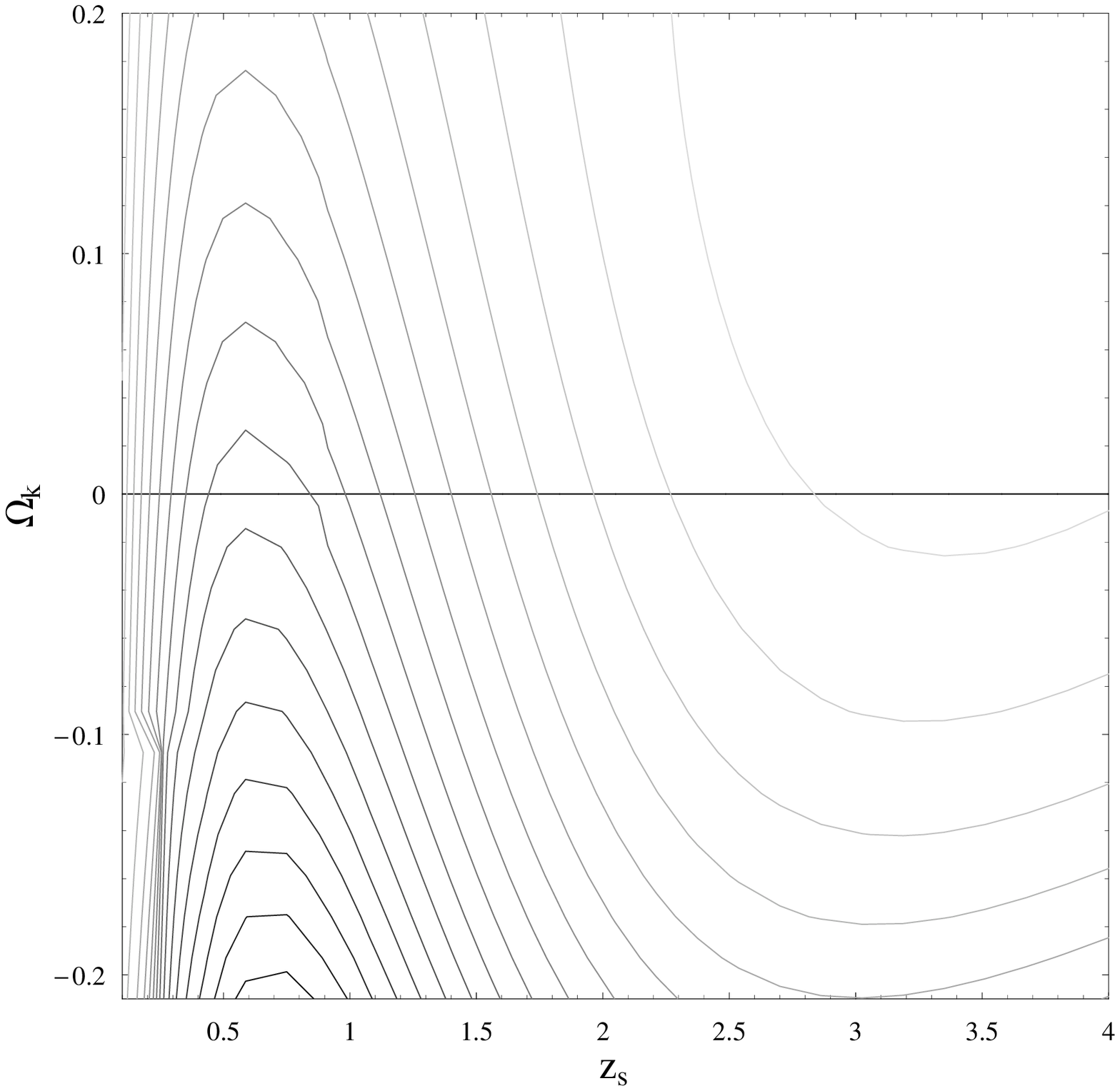}\label{fig14b}}
    \caption{.}\label{fig14}
\end{figure}

\begin{figure}[!hbtp]
  \centering
    \subfigure[ \hspace{0.1cm} Iso-$AP$ contours with $1\%$ error around
    $(\Omega_k=0,w_1=0)$. Solid lines are for $z=1.5,w_0=-0.4$,
    dashed lines for $z=0.5,w_0=-0.8$ and dot-dashed lines for $z=0.5,w_0=-2$.]
    {\includegraphics[width=0.4\textwidth]{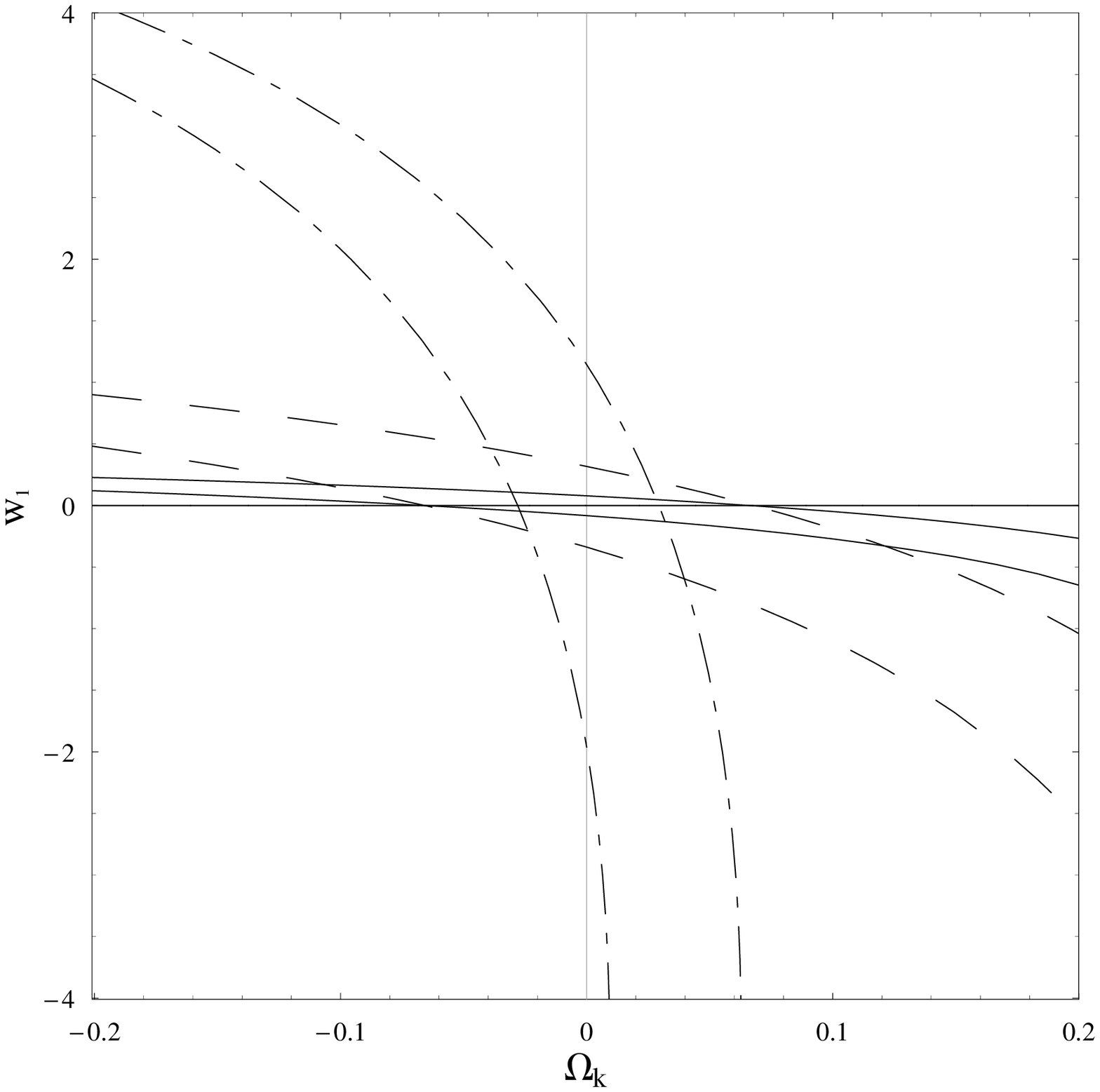}\label{fig15a}}
    \hspace{0.01\textwidth}
    \subfigure[ \hspace{0.1cm} Error sensitivity depending on $w_0$ and the redshift.]
    {\includegraphics[width=0.4\textwidth]{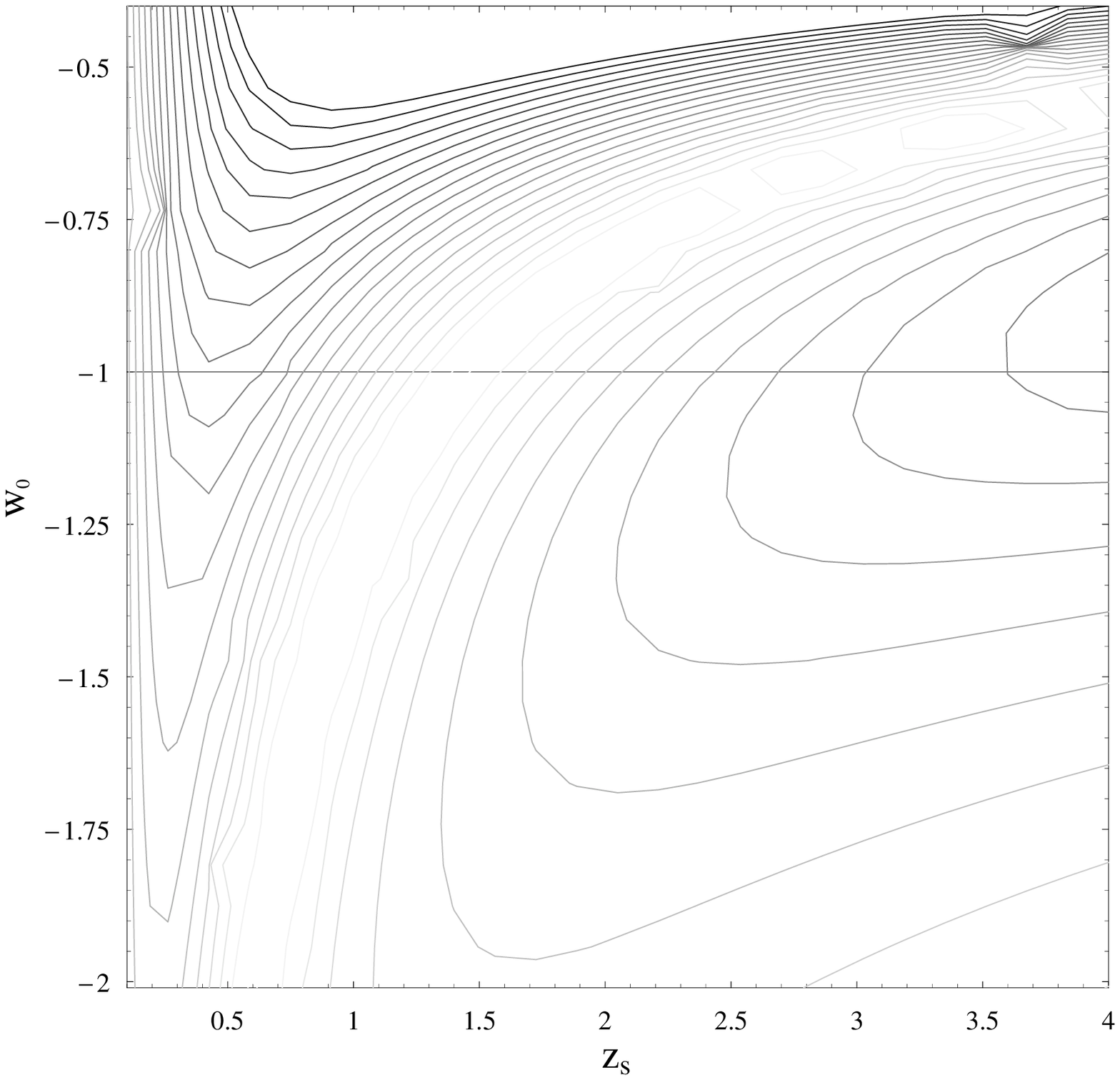}\label{fig15b}}
    \caption{.}\label{fig15}
\end{figure}

Studying the width of band for the same relative error of the
observable AP parameter, we obtained the possibility of obtaining
the tighter bound in the parameter spaces $(\Omega_k,w_0)$
(Fig.\ref{fig13}) and $(w_0, w_1)$ (Fig.\ref{fig14}) for the same
requirements as that of the observable $d_L$. In the parameter space
$(\Omega_k,w_1)$, the narrower band of AP for the same relative
error can be got at low redshift and $-1<w_0<-1/3$
(Fig.\ref{fig15}), which is not exactly the same as that of the
luminosity distance.

For the same relative error, a narrower band of an observable is
more interesting, since it can put more stringent constraints on the
parameter space, which can be more useful to reduce the cosmic
confusion if we employ the properties we learnt in
Fig.\ref{fig1}-\ref{fig9}.

In summary, we have extended the discussion of the uncertainty on
the determination of the DE of constant EoS due to a non-vanishing
spatial curvature by considering the luminosity distance \cite{11} to the dymanic DE models by considering
other fundamental observables. We discussed the sensitivity of these
observables to the value and redshift history of the EoS and the
spatial curvature, investigated whether these different observables
are complementary. Since the observables we studied in this paper are
measured nearly in all proposed tests of DE, we expect that our
analysis can be useful in the future observations in determining the
nature of the DE.

\begin{acknowledgments}
This work was partially supported by  NNSF of China, Ministry of
Education of China and Shanghai Education Commission. B. Wang would like to acknowledge helpful discussions with E. Abdalla and R. G. Cai.
\end{acknowledgments}


\end{document}